\documentclass[aps,prl,twocolumn,showpacs,preprintnumbers,float,longbibliography,superscriptaddress,notitlepage]{revtex4-2}

\usepackage[utf8]{inputenc}
\usepackage{epsfig}
\usepackage{float,amsmath,amssymb}
\usepackage{graphicx,yfonts}
\usepackage{url}
\usepackage{bbold}
\usepackage{physics}
\usepackage{enumitem}

\usepackage[]{mdframed}
\newcommand{\trento}{T$\mathrel{\protect\raisebox{-2.1pt}{R}}$ENTo}

\usepackage[breaklinks,colorlinks,citecolor=citcolor,urlcolor=blue,linkcolor=lcolor]{hyperref}
\definecolor{lcolor}{rgb}{0.5,0,0}
\definecolor{citcolor}{rgb}{0,0.3,0.0}

\begin{document}
	\title{Revealing Nuclear Shape through Identified-Hadron Flow Decorrelations}
	\begin{abstract}
		Relativistic nuclear collisions encode nuclear shape in collective flow, yet the potential of identified hadrons to disentangle deformation modes remains largely unexplored. Using event-by-event viscous hydrodynamics coupled to hadronic transport, we investigate momentum-dependent elliptic-flow decorrelations of pions, kaons, and protons in central $^{238}$U+$^{238}$U collisions at 193 GeV and $^{20}$Ne+$^{20}$Ne collisions at 5.36 TeV. The identified species exhibit distinct momentum- and deformation-dependent responses, with nuclear-shape information persisting in their relative correlations even when individual deformation effects are weak. We introduce crossing coordinates $(P_{\rm cross},R_{\rm cross})$, defined by the intersection of pion--kaon and pion--proton flow-factorization ratios. Variations of quadrupole deformation and triaxiality generate distinct trajectories in the crossing plane, offering complementary constraints on both nuclear-shape parameters. The corresponding behavior persists in higher-moment correlations, while an energy comparison in Ne+Ne reveals the influence of collision dynamics. Our findings establish identified-particle flow decorrelations as a complementary approach to probing and constraining intrinsic nuclear geometry.
	\end{abstract}

	\author{Hadi Mehrabpour}
	\email{mehrabpour@fudan.edu.cn}
	\affiliation{Institute of Modern Physics, Fudan University, Shanghai 200433, China}
	\author{Abhisek Saha}
	\email{saha@pku.edu.cn}
	\affiliation{School of Physics, Peking University, Beijing 100871, China}
    \affiliation{Center for High Energy Physics, Peking University, Beijing 100871, China}
	\author{Li Yan}
	\email{cliyan@fudan.edu.cn}
	\affiliation{Institute of Modern Physics, Fudan University, Shanghai 200433, China}
	\affiliation{Key Laboratory of Nuclear Physics and Ion-beam Application (MOE), Fudan University, Shanghai 200433, China}
	\affiliation{Shanghai Research Center for Theoretical Nuclear Physics, Shanghai 200438, China}

	\maketitle
	
	\textit{Introduction.--} Relativistic nuclear collisions probe nuclear structure through its imprint on the collective expansion of the quark--gluon plasma (QGP)~\cite{Jia:2022ozr,Giacalone:2025vxa,Duguet:2025qxi,Chen:2026gka}, with experimental constraints on nuclear shape from anisotropic and radial flow~\cite{STAR:2024wgy,STAR:2025elk}. Flow observables are sensitive to quadrupole, octupole, and hexadecapole deformation~\cite{Zhang:2021kxj,Mehrabpour:2023ign,Ryssens:2023fkv,Xu:2024bdh,3n5q-m2kf}, triaxiality and shape fluctuations~\cite{Jia:2021qyu,Zhao:2024lpc,Dimri:2023wup,Liu:2025fnq,Giacalone:2026fat,Bao:2026luj}, and nuclear density and collision dynamics~\cite{Giacalone:2017dud,Nijs:2021kvn,Fortier:2023xxy,Fortier:2024yxs,Mantysaari:2024uwn,Lu:2023fqd,Li:2026igf}. Mean transverse momentum, multiparticle correlations, and nonlinear flow provide complementary sensitivity~\cite{Xu:2021uar,Zhao:2022uhl,Nielsen:2023znu,Parida:2026uld,47jc-xzxm,Mehrabpour:2026lhj}. Studies of neutron skins and surface diffuseness~\cite{Giacalone:2023cet,Zhao:2026zno,Mehrabpour:2026dmu,Pihan:2025pep,Vitsos:2025jzt}, and of intrinsic shape and many-body correlations~\cite{Barata:2026aak,Blaizot:2025scr,Blaizot:2025bfu,Mehrabpour:2026yuc,Bofos:2026nmg}, highlight the need to distinguish structural effects from the dynamical response. Light-ion collisions extend this program to microscopic correlations and $\alpha$ clustering~\cite{Rybczynski:2019adt,Summerfield:2021oex,Zhang:2024vkh,Zhao:2024feh,Mehrabpour:2025ogw,Liu:2025uks,Lu:2025cni,Li:2025bdn,Ambrus:2026ptu,Duguet:2026jdf}, particularly in $^{20}$Ne~\cite{Giacalone:2024luz,Giacalone:2024ixe,Liu:2025zsi,Li:2025hae}. Related directions include longitudinal flow decorrelations~\cite{Mehrabpour:2025rzt,Garcia-Montero:2025bpn}, electron--ion probes~\cite{Magdy:2024thf}, and connections to nuclear matrix elements~\cite{Li:2025vdp}. Despite these advances, an important question remains: can identifying the produced hadrons reveal additional nuclear-shape information beyond that encoded in inclusive flow measurements?
	
	The opportunity arises because different hadron species need not carry identical imprints of the same initial nuclear geometry. The fluctuating collective flow vectors, $V_n=v_ne^{in\Psi_n}$~\cite{Voloshin:1994mz,Poskanzer:1998yz,Borghini:2000sa}, exhibit momentum-dependent factorization breaking~\cite{Gardim:2012im,Heinz:2013bua,Bozek:2018nne}, whose sensitivity to nuclear deformation has already been established~\cite{Bozek:2021mov,Samanta:2023qem}. Although pions, kaons, and protons emerge from a common evolving medium, their distinct masses, momentum-dependent collective responses, and hadronic interactions~\cite{Heinz:2013bua,Kisiel:2014upa} can encode the initial geometry differently in their flow correlations. Comparing identified-hadron decorrelations thus offers a new avenue to uncover nuclear-shape imprints that may be obscured in inclusive measurements and, more importantly, to disentangle deformation parameters that otherwise produce similar collective-flow signatures.
	
	Here, we investigate whether identified-particle flow correlations provide complementary constraints on quadrupole deformation and triaxiality using event-by-event viscous hydrodynamics coupled to hadronic transport. To systematically vary the initial nuclear geometry, we employ deformed Woods--Saxon densities~\cite{Miller:2007ri,PhysRev.95.577},
	\begin{align}
		&\rho(r,\theta,\phi)
		=\rho_0[1+\exp[(r-R(\theta,\phi))/a_0]]^{-1},\\
		&R(\theta,\phi)
		=R_0\bigl[1+\beta_2\cos\gamma\,Y_{20}(\theta,\phi)
		+\beta_2\sin\gamma\,Y_{22}^{c}(\theta,\phi)\bigr],
	\end{align}
	where \(\beta_2\geq0\) and \(0\leq\gamma\leq60^\circ\) characterize the quadrupole magnitude and triaxiality~\cite{Jia:2021qyu}, with \(Y_{22}^{c}=(Y_{22}+Y_{2,-2})/\sqrt{2}\). For \(^{20}\mathrm{Ne}\), this controlled shape scan isolates quadrupole deformation without explicitly incorporating microscopic cluster correlations.
	
	\begin{figure*}[t!]
		\hspace*{-.25cm}\includegraphics[scale=.24]{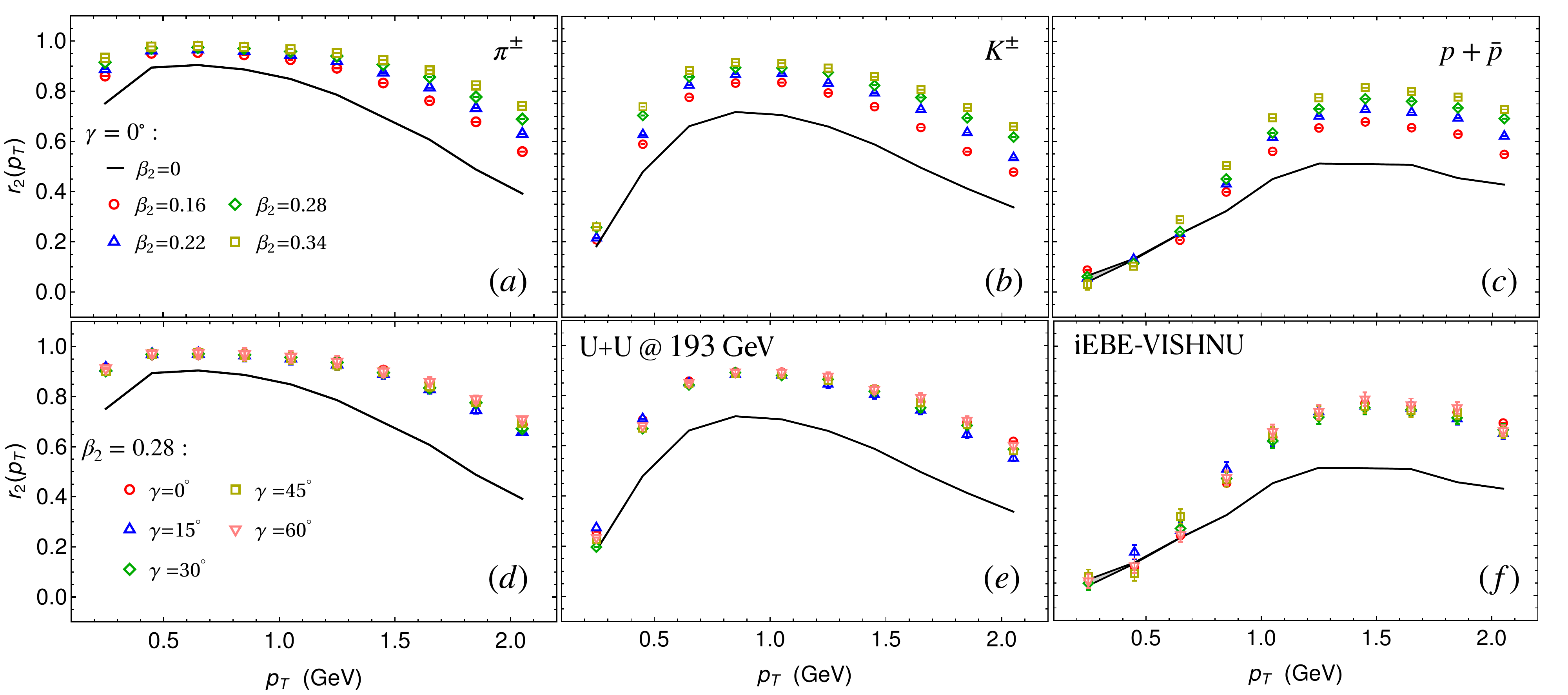}
		\caption{Transverse-momentum dependence of the identified-particle flow factorization coefficient $r_{2}(p_{\mathrm T})$ in central $^{238}$U+$^{238}$U collisions. Panels (a)--(c) show the evolution with quadrupole deformation $\beta_2$ for pions, kaons, and protons, respectively, while panels (d)--(f) present the corresponding dependence on the triaxiality angle $\gamma$ at fixed $\beta_2=0.28$. The identified hadrons exhibit distinct momentum-dependent correlations and pronounced responses to quadrupole deformation, whereas variations of triaxiality produce comparatively modest changes in the individual coefficients.}
		\label{fig:1}
	\end{figure*}
	
	For each identified species \(h=\pi^\pm,K^\pm,p+\bar p\), we quantify momentum-dependent flow correlations through the two-particle factorization coefficient
	\begin{equation}
		r_2^h(p_{\mathrm T})=
		\operatorname{Re}\left\langle
		V_2^{\rm ref}\,[V_2^h(p_{\mathrm T})]^*\right\rangle
		/[\left\langle (v_2^{\rm ref})^2\right\rangle
		\left\langle v_2^h(p_{\mathrm T})^2\right\rangle]^{\frac{1}{2}},
		\label{eq:r2}
	\end{equation}
	and its higher-moment counterpart~\cite{Bozek:2021mov,Samanta:2023qem},
	\begin{equation}
		r_{2,2}^h(p_{\mathrm T})=\operatorname{Re}
		\left\langle
			(V_2^{\rm ref})^2[V_2^h(p_{\mathrm T})^*]^2
			\right\rangle
		/[\left\langle (v_2^{\rm ref})^4\right\rangle
				\left\langle v_2^h(p_{\mathrm T})^4\right\rangle]^{\frac{1}{2}}
		\label{eq:r22}
	\end{equation}
	Here, brackets denote ensemble averages, and all species share a charged-particle reference with \(0.2\leq p_{\mathrm T}\leq3.0\) GeV. In the collective-flow limit, these moments can be estimated from distinct-particle azimuthal correlations~\footnote{Experimental application requires a consistent reference selection, self-correlation removal, and control of nonflow.}. We construct pion--kaon and pion--proton ratios of both factorization coefficients and introduce their crossing coordinates, defined by the intersection momentum and common ratio value. Their distinct responses to \(\beta_2\) and \(\gamma\) offer complementary constraints on quadrupole deformation and triaxiality. Our simulations combine \trento\, initial conditions~\cite{Moreland:2014oya}, event-by-event viscous hydrodynamics~\cite{Song:2010aq,Song:2007ux}, and UrQMD hadronic transport~\cite{Bass:1998ca,Bleicher:1999xi}. The hydrodynamic evolution begins at $\tau_0=0.6$ fm/$c$, with a specific shear viscosity $\eta/s=0.08$ and a switching temperature $T_{\rm sw}=160$ MeV. We study central \(^{238}\mathrm{U}+{}^{238}\mathrm{U}\) collisions at \(\sqrt{s_{\rm NN}}=193\) GeV (0--1\%) and \(^{20}\mathrm{Ne}+{}^{20}\mathrm{Ne}\) at 5.36 TeV (0--5\%), with identified particles within \(|\eta|<1\), using \(5\times10^4\) and \(10^5\) hydrodynamic events and 50 and 100 hadronic oversamples, respectively. Replacing UrQMD with SMASH~\cite{SMASH:2016zqf,Petersen:2018jag} introduces quantitative differences but preserves the qualitative deformation-dependent patterns, indicating that the nuclear-structure imprints are not specific to the choice of hadronic afterburner.
	
	\textit{Species-dependent flow decorrelations.—}
	Figure~\ref{fig:1} illustrates $r_2(p_{\mathrm T})$ for central U+U collisions. Even for spherical nuclei, the coefficients exhibit distinct momentum dependences: pion correlations peak at low $p_{\mathrm T}$, kaons at intermediate momentum, and protons exhibit a broader maximum at higher $p_{\mathrm T}$. These profiles reflect the different momentum-dependent responses of identified hadrons to the common collective expansion, influenced by their masses and hadronic interactions. At $p_{\mathrm T}\simeq2.1$~GeV, the spherical coefficients decrease to approximately $0.4$, $0.35$, and $0.45$ for pions, kaons, and protons, respectively. Increasing the quadrupole deformation to $\beta_2=0.34$ at $\gamma=0^\circ$ raises them to approximately $0.72$, $0.65$, and $0.71$, indicating weaker flow decorrelations. At $p_{\mathrm T}\simeq1.7$~GeV, the pion coefficient increases from $\sim0.60$ to $\sim0.84$ and the kaon coefficient from $\sim0.50$ to $\sim0.78$, while protons exhibit a weaker response in this momentum region. These differences demonstrate that quadrupole deformation modifies the momentum-dependent correlations of identified hadrons non-universally.
	
	At fixed $\beta_2=0.28$, variations of the triaxiality angle $\gamma$ produce considerably smaller changes, with substantial overlap among the configurations for all three species. Even at $p_{\mathrm T}\simeq2.1$~GeV, the proton coefficients remain closely grouped around $0.65$--$0.70$. Thus, while quadrupole deformation produces a pronounced enhancement of the normalized flow correlations, the triaxial response is comparatively weak in the individual coefficients. The distinct momentum profiles and deformation dependences nevertheless demonstrate that identified hadrons encode different aspects of the collective response to the initial nuclear geometry.
	
\begin{figure}[t!]
		\hspace*{-.2cm}\includegraphics[width=0.45\textwidth]{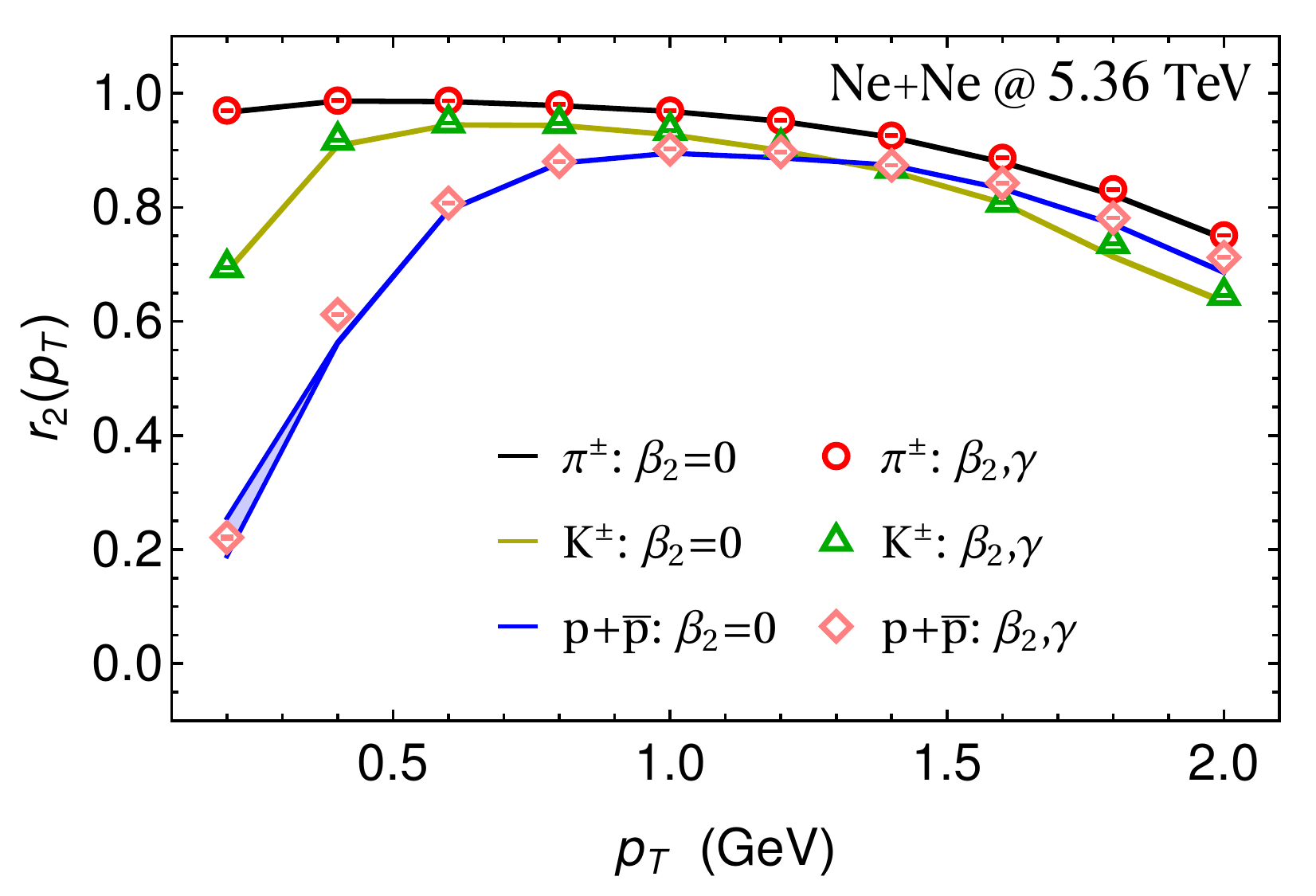}
		\caption{Identified-particle $r_2(p_T)$ in central $^{20}$Ne+$^{20}$Ne collisions at 5.36~TeV (0--5\%), comparing spherical nuclei with $(\beta_2,\gamma)=(0.4,15^\circ)$. All three species are shown in the same panel.}
		\label{fig:2}
	\end{figure}

	Figure~\ref{fig:2} shows $r_2(p_{\mathrm T})$ for central $^{20}$Ne+$^{20}$Ne collisions at $\sqrt{s_{\rm NN}}=5.36$~TeV, comparing spherical and triaxially deformed $(\beta_2,\gamma)=(0.4,15^\circ)$ nuclei~\footnote{Microscopic approaches describe many-body correlations, collective deformation, and clustering~\cite{Sun:2026yrr,Lonardoni:2018nob,Frosini:2021ddm,Elhatisari:2017eno,Meissner:2014lgi}. Such configurations have been used in relativistic collisions~\cite{Summerfield:2021oex,Zhang:2024vkh,Mehrabpour:2025ogw,Liu:2025uks,Liu:2025zsi,Li:2025hae}. Our WS scan varies quadrupole shape without explicit cluster correlations. Charge radii~\cite{ANGELI201369} constrain size but do not uniquely determine the matter density.}. Unlike spherical U+U collisions, the Ne+Ne coefficients remain strongly correlated with the reference flow over a broader momentum range. Pion correlations stay close to unity up to $p_{\mathrm T}\sim1$~GeV, while kaon and proton coefficients rise from approximately $0.70$ and $0.22$ at $p_{\mathrm T}\simeq0.2$~GeV to maxima near $0.95$ and $0.90$, respectively. The kaon and proton coefficients approach each other around $p_{\mathrm T}\sim1.3$~GeV, with protons becoming more strongly correlated at higher momentum. At $p_{\mathrm T}\simeq2$~GeV, the spherical coefficients remain around $0.75$, $0.65$, and $0.70$ for pions, kaons, and protons, respectively. The deformation-induced differences are much weaker than in U+U, remaining at the $\sim1$--$2\%$, $\sim2$--$4\%$, and few-percent levels, respectively. The higher-moment results in the Supplemental Material (Figs.~7 and 15(a)) reveal a further collision-energy dependence: lowering $\sqrt{s_{\rm NN}}$ from $5.36$~TeV to $200$~GeV modifies the species-dependent momentum profiles and generally enhances decorrelations, with $r_{2,2}^h(200~\mathrm{GeV})/r_{2,2}^h(5.36~\mathrm{TeV})\sim0.8$--$0.9$ near $p_{\mathrm T}\simeq2$~GeV. These results highlight the dependence of identified-hadron correlations on the collision dynamics, while their distinct fractional deformation responses motivate constructing species ratios that suppress common modifications and retain complementary nuclear-shape information.
	
	\textit{Species-dependent response to nuclear deformation.—}
	To characterize the relative deformation sensitivity of
	identified-particle flow correlations, we introduce the
	species-response ratio
	\begin{equation}\label{eqR}
		R_{\pi,h}^{(1)}(p_{\mathrm T})
		=
		\frac{r_2^\pi(p_{\mathrm T})}
		{r_2^h(p_{\mathrm T})}
		=
		\sqrt{
			\frac{\left\langle v_2^2(p_{\mathrm T})\right\rangle_h}
			{\left\langle v_2^2(p_{\mathrm T})\right\rangle_\pi}
		}
		\frac{
			\operatorname{Re}\left\langle
			V_2V_2^*(p_{\mathrm T})\right\rangle_\pi
		}{
			\operatorname{Re}\left\langle
			V_2V_2^*(p_{\mathrm T})\right\rangle_h
		},
	\end{equation}
	where $h=K^{\pm}$, $p+\bar{p}$.
	The second equality assumes a common charged-particle
	reference vector and a consistent ensemble-averaging
	prescription for both species.
	A common multiplicative modification cancels,
	whereas unequal fractional changes of the two
	coefficients modify $R_{\pi,h}^{(1)}$.
	Its value need not be unity even for spherical nuclei; nuclear-shape sensitivity is therefore characterized by its variation between deformation configurations.
	
\begin{figure}[t!]
		\includegraphics[width=0.45\textwidth]{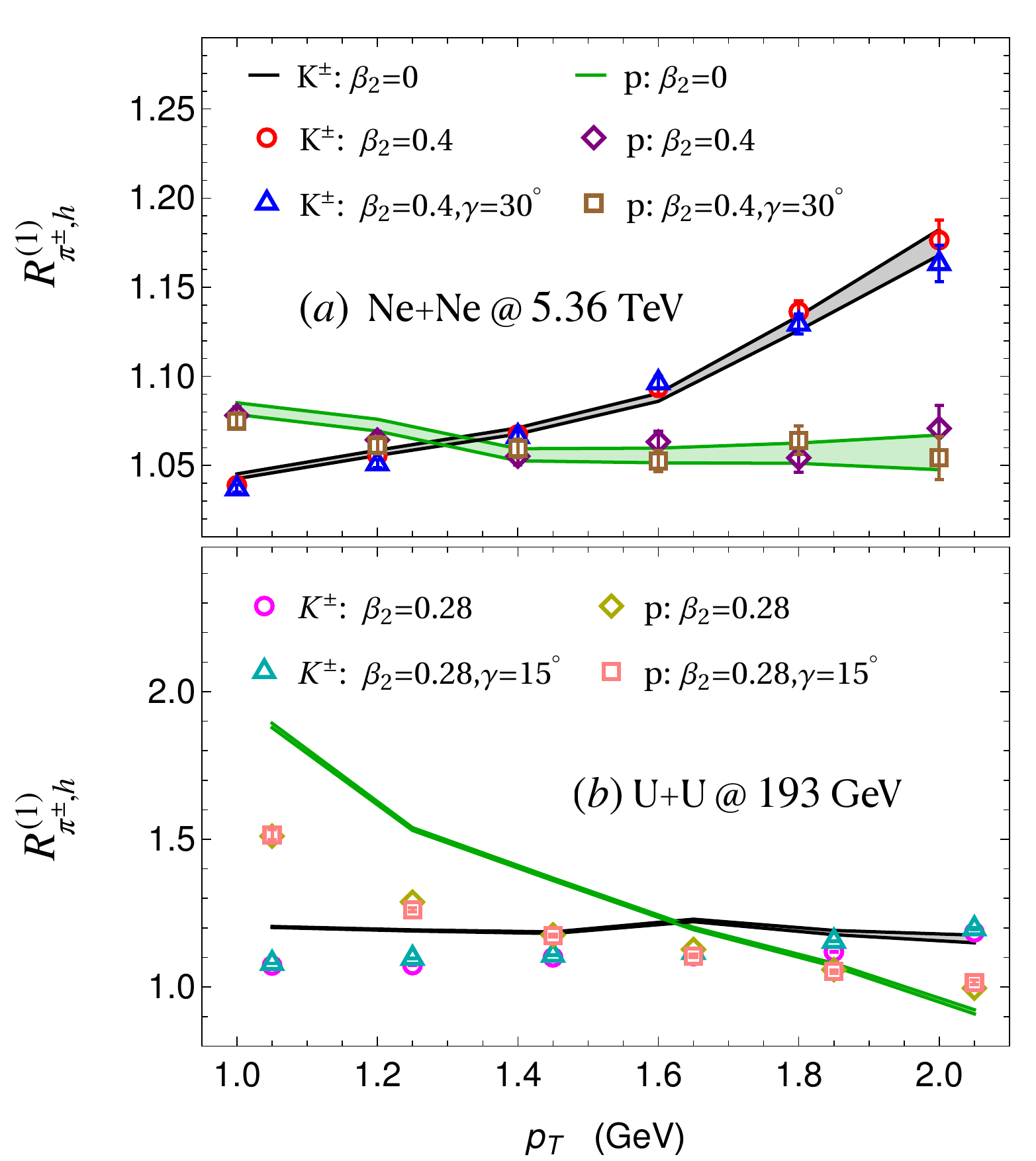}
		\caption{Species ratios $R_{\pi,K}^{(1)}$ and $R_{\pi,p}^{(1)}$ for spherical, axial, and triaxial configurations: (a) Ne+Ne at 5.36~TeV (0--5\%); (b) U+U at 193~GeV (0--1\%).}
		\label{fig:4}
	\end{figure}

	Figure~\ref{fig:4} compares
	$R_{\pi^\pm,K^\pm}^{(1)}$ and
	$R_{\pi^\pm,p}^{(1)}$ for spherical, axially
	symmetric, and triaxially deformed configurations.
	In Ne+Ne collisions [panel~(a)], both ratios remain
	close to unity for $p_{\mathrm T}\lesssim1.5$~GeV,
	with only weak deformation dependence.
	At higher momentum, their momentum dependences
	separate: for $\beta_2=0.4$,
	$R_{\pi^\pm,K^\pm}^{(1)}$ increases markedly with
	$p_{\mathrm T}$, whereas
	$R_{\pi^\pm,p}^{(1)}$ remains nearly flat around
	$1.05$--$1.08$.
	Introducing triaxiality with $\gamma=30^\circ$
	produces a modest additional modulation at high
	$p_{\mathrm T}$.
	The separation of the two ratios reflects differences
	among the species coefficients, while the changes
	between nuclear configurations characterize their
	relative deformation sensitivity.
	
	The configuration dependence is more pronounced
	in U+U collisions [panel~(b)].
	For spherical nuclei,
	$R_{\pi^\pm,p}^{(1)}$ exhibits a strong enhancement
	at the lower end of the displayed interval,
	reaching $\sim1.9$ around
	$p_{\mathrm T}\simeq1$~GeV before decreasing toward
	$p_{\mathrm T}\simeq2$~GeV.
	Quadrupole deformation with $\beta_2=0.28$
	substantially suppresses this enhancement,
	while triaxiality with $\gamma=15^\circ$
	further modulates both species ratios across
	the momentum range.
	Changing nuclear shape therefore modifies the relative momentum dependence, including cases with weak individual-species changes. These results show that quadrupole deformation and triaxiality modify the relative momentum dependence of the identified-particle responses.
	
	\begin{figure*}[t!]
		\hspace*{-.2cm}\includegraphics[scale=.26]{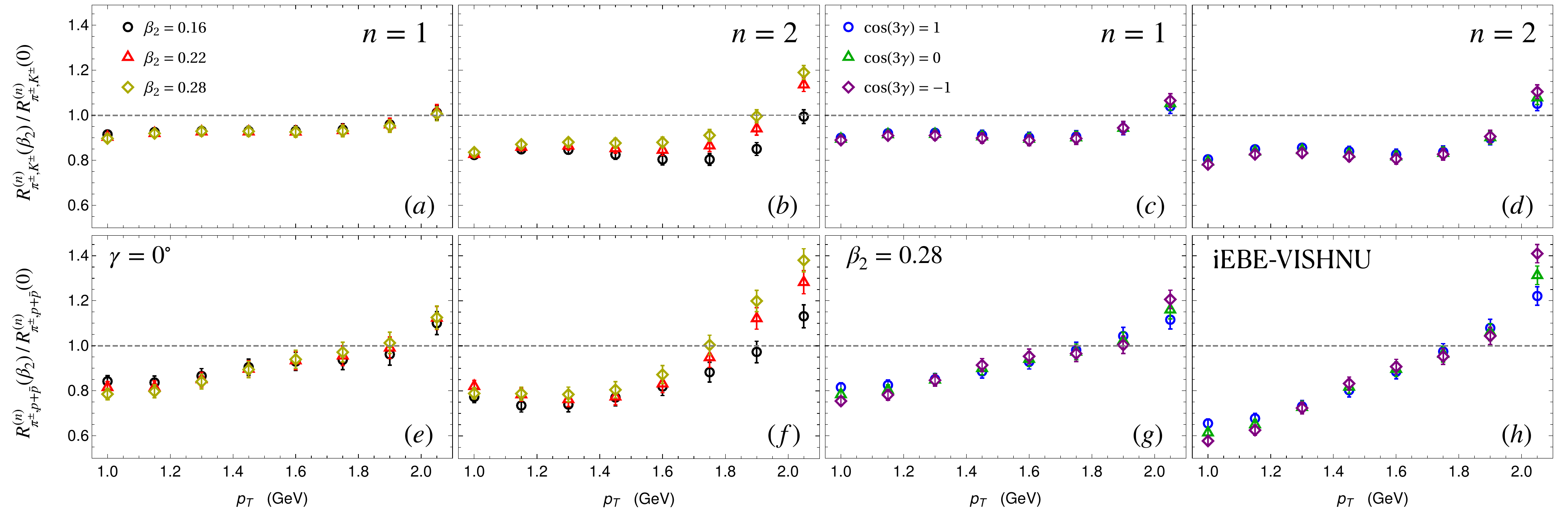}
		\caption{Normalized species ratios $\mathcal R_{\pi,h}^{(n)}$ in central U+U collisions. Top row: pion--kaon; bottom row: pion--proton. Columns 1--2 scan $\beta_2$ at $\gamma=0^\circ$; columns 3--4 scan $\gamma$ at $\beta_2=0.28$. Within each pair, $n=1$ ($r_2$) precedes $n=2$ ($r_{2,2}$).}
		\label{fig:3}
	\end{figure*}
	
	\textit{Deformation dependence of species ratios.—}
	To characterize the deformation-induced changes of the
	species ratios, we normalize each ratio to its spherical
	model baseline,
	\(\mathcal R_{\pi,h}^{(n)}\equiv R_{\pi,h}^{(n)}(\beta_2,\gamma;p_T)/R_{\pi,h}^{(n)}(0;p_T)\),
	where $n=1,2$ denotes the constructions based on
	$r_2$ and $r_{2,2}$, respectively.
	This normalization quantifies the relative change
	between deformed and spherical calculations; the
	spherical baseline is a model reference.
	For moderate deformations, we parameterize this
	dependence as
	\begin{align}
		\frac{R_{\pi,h}(\beta_2,\gamma)}
		{R_{\pi,h}(0)}
		\simeq
		1
		&+\delta A_{\pi,h}(p_T)\beta_2^2
		+\delta B_{\pi,h}(p_T)\beta_2^3\cos(3\gamma)
		\nonumber\\
		&+\delta C_{\pi,h}^{\rm eff}(p_T)\beta_2^4,
		\label{eq:response}
	\end{align}
	where the coefficients describe the momentum-dependent
	deformation response of the species ratio.
	The quartic coefficient includes the contribution
	generated by expanding the denominator of the ratio.
	The coefficient extraction and the corresponding
	decompositions for both constructions are presented
	in the Supplemental Material, see
	Figs.~\ref{fig:7} and \ref{fig:14}.
	
	Figure~\ref{fig:3} shows that the normalized response
	depends on both species and momentum.
	For $\gamma=0^\circ$ and $\beta_2=0.28$, the normalized
	pion--kaon ratio constructed from $r_{2,2}$ increases
	from about $0.83$ at $p_T\simeq1$~GeV to $1.19$
	at $p_T\simeq2.05$~GeV, while the pion--proton ratio
	changes from about $0.78$ to $1.37$ over the same
	range.
	The corresponding $r_2$ ratios exhibit the same
	qualitative momentum dependence, with smaller
	deformation-induced variations.
	Thus, the fractional changes of the pion and
	comparison-species coefficients need not have the
	same ordering throughout the momentum interval:
	a normalized ratio below unity indicates a smaller
	fractional change for pions, whereas a value above
	unity indicates a larger one.
	
	Within the parameterization, the decomposition
	in Fig.~\ref{fig:7}(a),(b) exhibits substantial
	cancellations among the quadratic, cubic, and
	quartic contributions, with the quartic term
	remaining non-negligible.
	At fixed $\beta_2=0.28$, the $\gamma$ dependence
	is weaker for the pion--kaon ratio but becomes
	increasingly pronounced for the pion--proton ratio
	toward higher $p_T$, as seen in
	Fig.~\ref{fig:3}(c),(d) and (g),(h).
	To the order retained in Eq.~\eqref{eq:response},
	the angular dependence enters through
	$\beta_2^3\cos(3\gamma)$, while the quadratic and
	quartic terms are independent of $\gamma$.
	The corresponding contributions are shown in
	Fig.~\ref{fig:7}(c)--(f).
	Both correlator constructions therefore exhibit
	species-dependent deformation responses, with
	larger variations in the four-particle construction.
	These complementary momentum dependences motivate
	the crossing observable introduced below.
	
	\begin{figure*}[t!]
		\hspace*{-.2cm}\includegraphics[scale=.27]{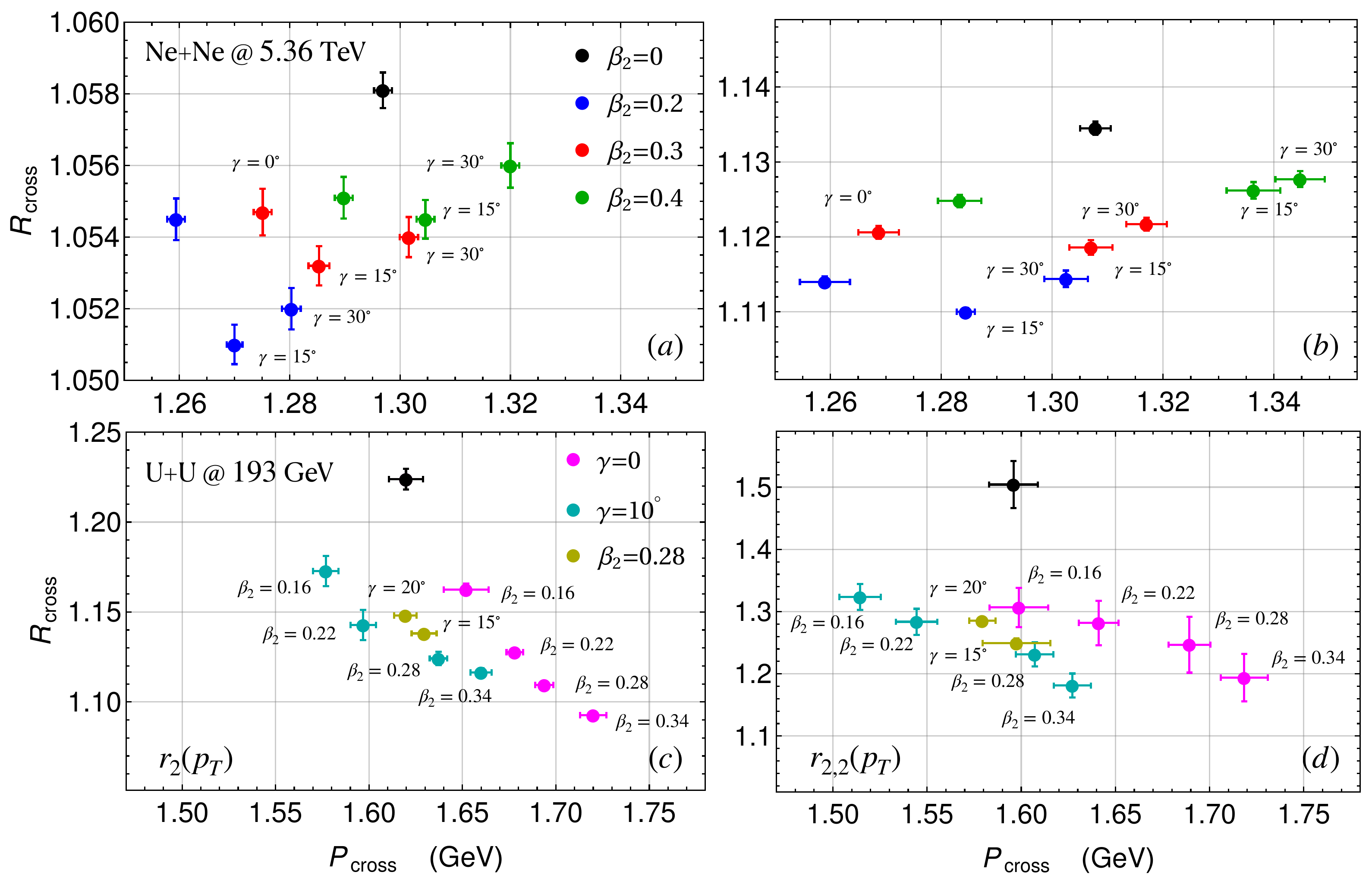}
		\caption{Crossing-coordinate maps \((P_{\rm cross},R_{\rm cross})\) constructed from the direct pion--kaon and pion--proton ratios of identified-particle flow-decorrelation coefficients in central $^{20}$Ne+$^{20}$Ne collisions at 5.36~TeV (0--5\%, upper panels) and $^{238}$U+$^{238}$U collisions at 193~GeV (0--1\%, lower panels). Panels (a),(c) correspond to $r_2$, while (b),(d) use $r_{2,2}$. Each point represents a nuclear configuration specified by $(\beta_2,\gamma)$, with crossing coordinates satisfying $R_{\pi,K}(P_{\rm cross})=R_{\pi,p}(P_{\rm cross})$. Horizontal and vertical error bars indicate uncertainties in $P_{\rm cross}$ and $R_{\rm cross}$, respectively. Note the different axis scales across collision systems and correlator constructions, particularly the magnified vertical scales in Ne+Ne, which resolve small deformation-dependent variations that would otherwise be difficult to distinguish.}
		\label{fig:5}
	\end{figure*}
	
	\textit{Crossing coordinates.—}
	To capture the complementary nuclear-shape information encoded in the momentum dependence of identified-particle ratios, we introduce the crossing coordinates
	\begin{equation}
		C_{\rm cross}\equiv(P_{\rm cross},R_{\rm cross}),
	\end{equation}
	defined by
	\begin{equation}
		R_{\pi,K}(P_{\rm cross})=R_{\pi,p}(P_{\rm cross})
		\equiv R_{\rm cross}.
	\end{equation}
	Since the equality of these ratios implies $r_2^K(P_{\rm cross})=r_2^p(P_{\rm cross})$, the crossing momentum $P_{\rm cross}$ identifies where kaon and proton correlations coincide, while $R_{\rm cross}$ measures the pion correlation relative to their common value. Rather than examining the full momentum dependence of each ratio separately, these two coordinates capture their relative momentum evolution and correlation strength in a compact form. Their joint variation can therefore provide complementary constraints on quadrupole deformation and triaxiality. The construction applies to both $r_2$ and $r_{2,2}$, with the corresponding index suppressed below.
	
	Figure~\ref{fig:5} shows that different combinations of $\beta_2$ and $\gamma$ generate distinct trajectories in the crossing plane. For Ne+Ne, the $r_2$ crossings occupy a narrow region, $P_{\rm cross}\simeq1.26$--$1.32$~GeV and $R_{\rm cross}\simeq1.051$--$1.058$, while $r_{2,2}$ expands the vertical range to $R_{\rm cross}\simeq1.110$--$1.135$. In U+U, the variations are considerably larger: $P_{\rm cross}\simeq1.58$--$1.74$~GeV and $R_{\rm cross}\simeq1.09$--$1.23$ for $r_2$, extending to $R_{\rm cross}\simeq1.18$--$1.50$ for $r_{2,2}$. The larger variation of the higher-moment construction provides additional sensitivity to nuclear shape. Accounting for the full momentum--height covariance, nearby deformation scenarios exhibit a two-sided Gaussian-equivalent separation of $Z\simeq2$--$5$ in our simulations~\footnote{For $\Delta C=C_{{\rm cross},1}-C_{{\rm cross},2}$, $D^2=\Delta C^T\Sigma_\Delta^{-1}\Delta C$ includes the momentum--height covariance; independent ensembles give $\Sigma_\Delta=\Sigma_1+\Sigma_2$. Under the equal-mean null, Gaussian fluctuations, and the known-covariance approximation, $D^2\sim\chi_2^2$, so $p=e^{-D^2/2}=2[1-\Phi(Z)]$, where $\Phi$ is the standard normal cumulative distribution function.}. Crucially, variations of $\beta_2$ and $\gamma$ produce different displacements in the crossing plane, rather than merely changing the correlation strength. This distinct two-dimensional response offers a complementary avenue for disentangling quadrupole deformation and triaxiality through identified-particle flow correlations.

	\textit{Summary.—}
	We have demonstrated a new approach to probing nuclear shape through identified-particle flow decorrelations in relativistic collisions. Using event-by-event viscous hydrodynamics coupled to hadronic transport, we studied central $^{238}$U+$^{238}$U and $^{20}$Ne+$^{20}$Ne collisions and found distinct momentum-dependent responses of pions, kaons, and protons to quadrupole deformation and triaxiality. Even when the deformation dependence of individual coefficients is weak, as in Ne+Ne, their relative correlations retain nuclear-shape information. Species ratios constructed from $r_2(p_{\mathrm T})$ and its higher-moment counterpart $r_{2,2}(p_{\mathrm T})$ reveal complementary deformation responses, with the latter exhibiting larger variations.
	
	Our principal result is the crossing-coordinate observable $(P_{\rm cross},R_{\rm cross})$, which captures the intersection of pion--kaon and pion--proton ratios. Different combinations of $(\beta_2,\gamma)$ generate distinct trajectories in the crossing plane, with nearby configurations typically separated at $Z\simeq2$--$5$ in our simulations. These two coordinates offer a compact, experimentally accessible means of disentangling quadrupole deformation and triaxiality. Although demonstrated here for quadrupole shapes, the framework can be extended to higher-order deformations, including octupole and hexadecapole components. Our findings motivate identified-particle flow-decorrelation measurements at RHIC and the LHC as complementary probes of intrinsic nuclear geometry.
	
	\textit{Acknowledgments.—}  We thank Piotr Bozek and Huichao Song for useful discussions. This work is partially supported by the National Natural Science Foundation of China under Grant numbers 12347106, 12375133 and 12147101.

	\bibliography{biblio, biblioNotes}
	
	\appendix
	\begin{figure*}[t!]
		\hspace*{-.25cm}\includegraphics[scale=.285]{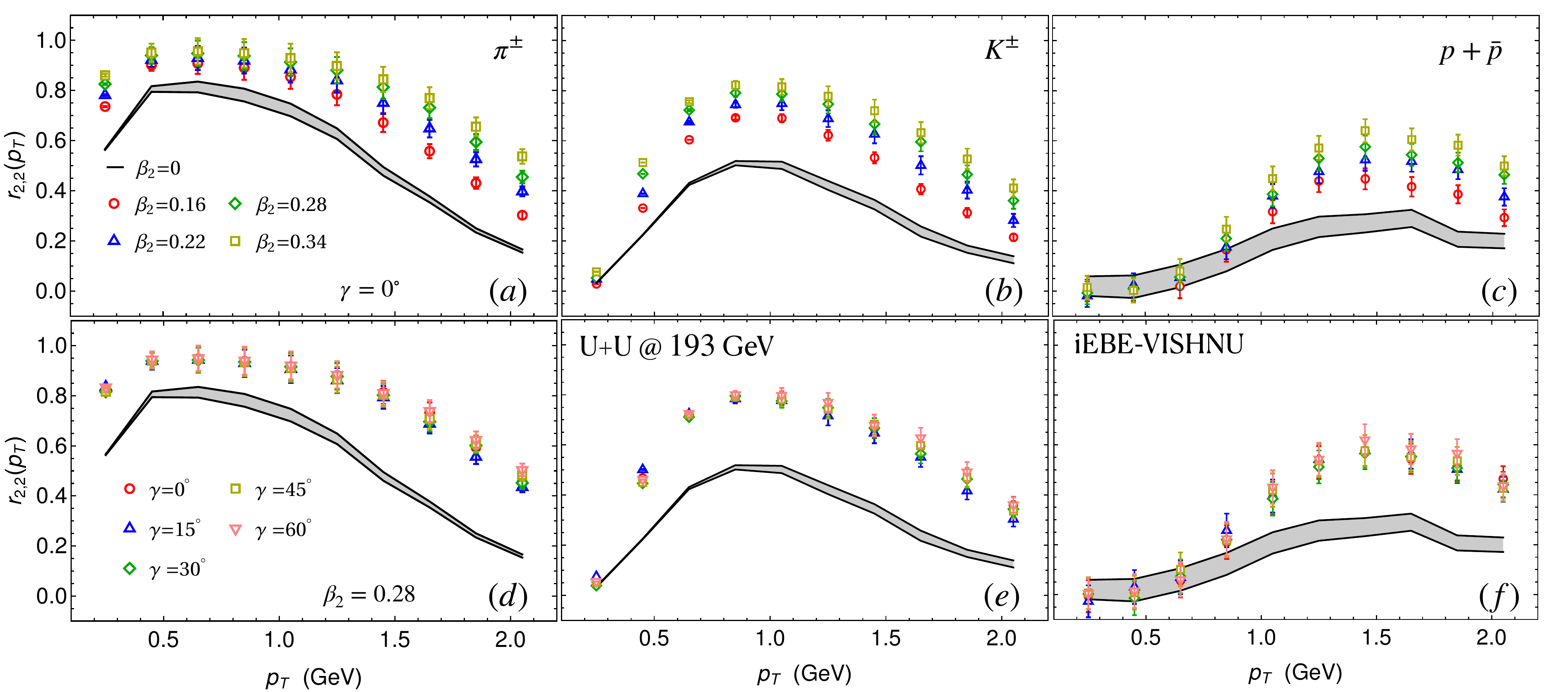}
		\caption{The same as Fig.~\ref{fig:1} but for higher-moment coefficient $r_{2,2}(p_T)$, defined in Eq.~\ref{eq:r22} and evaluated using he approximation in Eq.~\ref{eq:four}.}
		\label{fig:1-1}
	\end{figure*}
	\section*{Supplementary Materials}
	
	\subsection{Four-particle flow decorrelations}
	\label{sec:app_r22}
	
	As a higher-moment cross-check of the two-particle analysis in
	Figs.~\ref{fig:1} and~\ref{fig:2}, we consider the
	identified-particle flow decorrelation $r_{2,2}^h(p_{\mathrm T})$,
	defined in Eq.~\eqref{eq:r22}
	~\cite{Bozek:2021mov,Bozek:2022slu,Samanta:2023qem}.
	Its exact normalization involves the fourth moment of the
	differential flow, which is statistically challenging to
	determine, particularly for protons. We therefore employ
	the approximation
	\begin{equation}
		\left\langle [v_2^h(p_{\mathrm T})]^4\right\rangle
		\simeq
		\left\langle [v_2^h(p_{\mathrm T})]^2\right\rangle^2
		\frac{\left\langle (v_2^{\rm ref})^4\right\rangle}
		{\left\langle (v_2^{\rm ref})^2\right\rangle^2},
	\end{equation}
	which yields
	\begin{equation}\label{eq:four}
		r_{2,2}^h(p_{\mathrm T})
		\simeq
		\frac{
			\left\langle (v_2^{\rm ref})^2\right\rangle
			\left\langle (V_2^{\rm ref})^2
			[V_2^h(p_{\mathrm T})^*]^2\right\rangle
		}{
			\left\langle (v_2^{\rm ref})^4\right\rangle
			\left\langle [v_2^h(p_{\mathrm T})]^2\right\rangle
		}.
	\end{equation}
	This approximation assumes equal normalized fourth moments
	for the differential and reference flow distributions.
	The corresponding results for central $^{238}$U+$^{238}$U collisions are
	shown in Fig.~\ref{fig:1-1}. As in the two-particle observable,
	$r_{2,2}(p_{\mathrm T})$ exhibits a clear particle dependence and decreases
	toward high $p_{\mathrm T}$. Increasing the quadrupole deformation at
	$\gamma=0^\circ$ enhances the decorrelation coefficient for all three
	species, with the separation between deformation configurations becoming
	more pronounced at intermediate and high $p_{\mathrm T}$. At fixed
	$\beta_2=0.28$, variations of $\gamma$ produce a weaker modification of the
	individual responses, but the proton distribution develops a comparatively
	stronger separation at high $p_{\mathrm T}$. Thus, the four-particle
	correlator preserves the particle-dependent sensitivity to quadrupole
	deformation and triaxiality observed with $r_2(p_{\mathrm T})$.
	
\begin{figure}[t!]
		\hspace*{-.2cm}\includegraphics[width=0.5\textwidth]{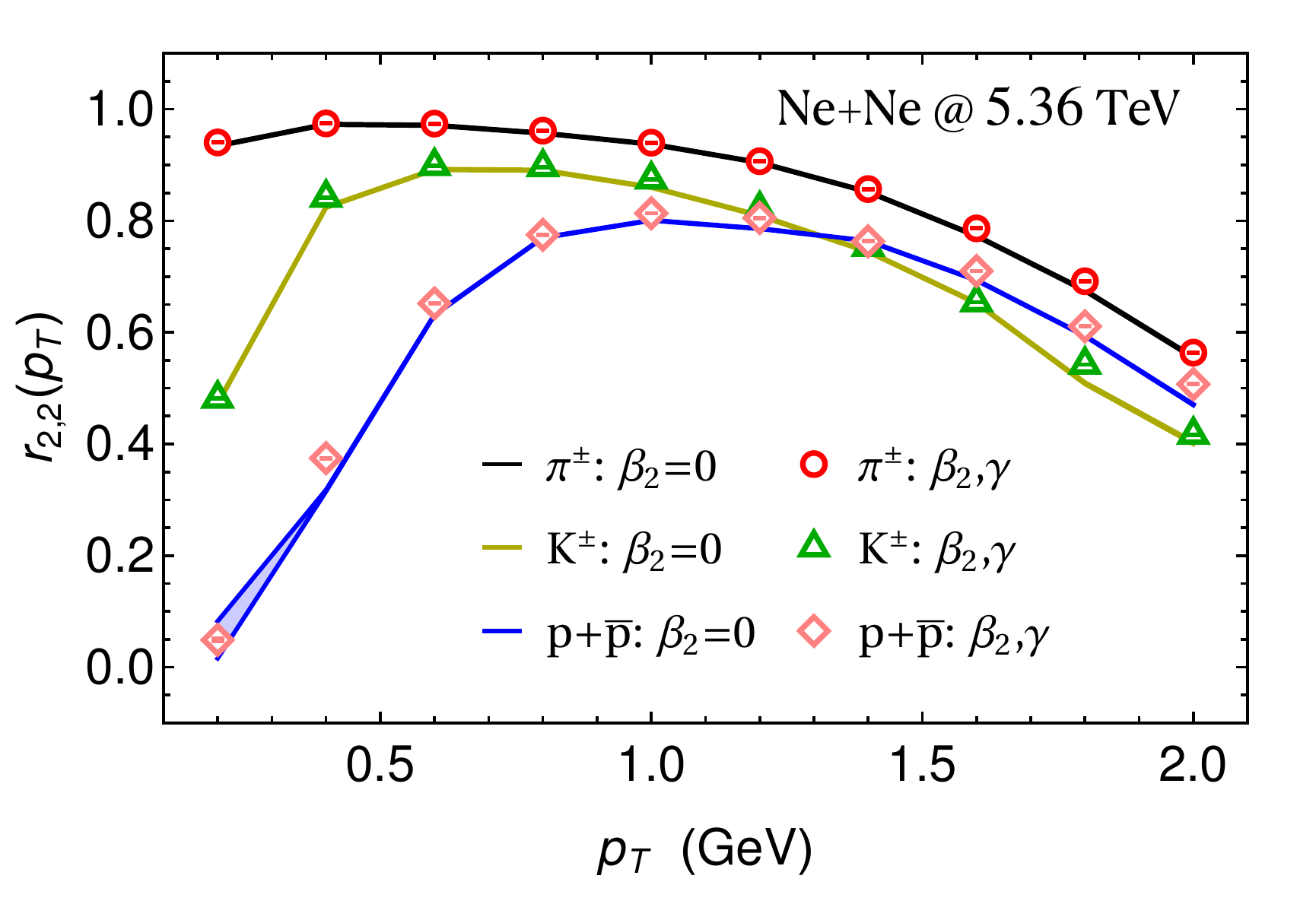}
		\caption{The same as Fig.~\ref{fig:2} but for $r_{2,2}(p_T)$.
		}
		\label{fig:2-2}
	\end{figure}

	The corresponding results for central $^{20}$Ne+$^{20}$Ne collisions are
	shown in Fig.~\ref{fig:2-2}. Comparing spherical nuclei with the
	triaxially deformed configuration $(\beta_2,\gamma)=(0.4,15^\circ)$ again
	shows only a small deformation dependence of the individual identified
	particles. The pion, kaon, and proton responses remain close to one another
	over most of the displayed momentum range, consistent with the suppression
	of the individual-particle deformation signal observed in Fig.~\ref{fig:2}.
	Nevertheless, the same qualitative particle-dependent response patterns
	are retained in $r_{2,2}(p_{\mathrm T})$.
	
	The agreement between the two- and four-particle observables demonstrates
	that the particle-dependent response to the initial nuclear geometry is
	not specific to the choice of correlator. The four-particle observable
	provides enhanced sensitivity to deformation, while the two-particle
	correlator is experimentally simpler and is therefore used as the primary
	observable in the main text.

	\begin{figure*}[t!]
		\includegraphics[scale=.27]{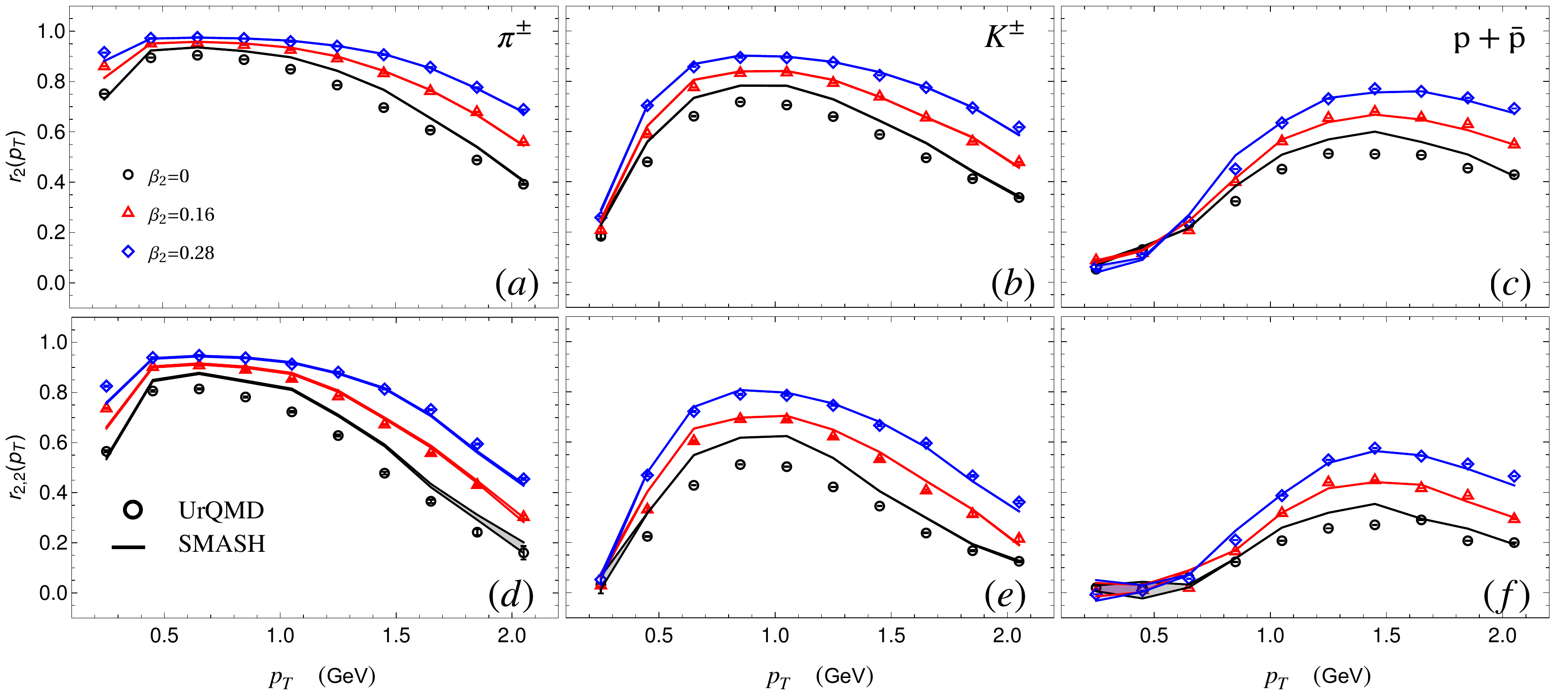}
		\caption{Sensitivity of the identified-particle flow decorrelations to
			the hadronic transport afterburner in central
			$^{238}$U+$^{238}$U collisions at
			$\sqrt{s_{\rm NN}}=193$~GeV. Panels (a)--(c) show $r_2(p_T)$ for
			pions, kaons, and protons, respectively, while panels (d)--(f) show the
			corresponding four-particle decorrelation $r_{2,2}(p_T)$. Open symbols
			denote results obtained with UrQMD and solid curves those obtained with
			SMASH for spherical ($\beta_2=0$) and deformed
			($\beta_2=0.16$ and $0.28$) configurations at $\gamma=0^\circ$.
			All other stages of the simulation, model inputs, and particle-selection
			cuts are kept fixed. The two hadronic transport descriptions produce
			quantitative differences, particularly for kaons and protons, while
			preserving the characteristic deformation ordering and its
			particle-dependent momentum dependence.
		}
		\label{fig:afterburner}
	\end{figure*}
	\textit{Hadronic-transport dependence.--}
	To assess the sensitivity of identified-particle correlations
	to the late hadronic evolution, we replace the UrQMD
	afterburner with SMASH while keeping the initial conditions,
	hydrodynamic evolution, particlization, and particle-selection
	cuts unchanged. UrQMD and SMASH provide alternative microscopic descriptions of the nonequilibrium hadronic stage, with differences in their transport implementations discussed in Ref.~\cite{Reichert:2021ljd}. SMASH is employed in its \texttt{List} mode, using externally sampled
	particles from hydrodynamics, with a fixed time step
	$\Delta t=0.1$~fm.
	
	Figure~\ref{fig:afterburner} compares $r_2(p_{\mathrm T})$
	[panels (a)--(c)] and $r_{2,2}(p_{\mathrm T})$
	[panels (d)--(f)] for pions, kaons, and protons in central
	U+U collisions. Open symbols represent UrQMD, while solid
	curves correspond to SMASH for identical nuclear
	configurations. Although the two afterburners produce
	quantitative differences, particularly for kaons and protons,
	the deformation-induced trends remain qualitatively
	unchanged. Both calculations preserve the characteristic
	ordering with $\beta_2$ over most of the momentum range
	and the distinct momentum dependence of the identified
	hadrons. The persistence of these patterns indicates that
	the nuclear-shape sensitivity is not specific to the hadronic
	transport prescription, while the quantitative differences
	reflect the contribution of late hadronic evolution to the
	final-state correlations.
	\subsection*{Competition between identified-particle response functions}
	The particle-response ratio introduced in the main text is designed to isolate how different identified particles encode the same initial nuclear geometry after the hydrodynamic evolution. While the two-particle ($r_{2}$) and four-particle ($r_{2,2}$) decorrelations measure the collective response of identified hadrons, they contain two qualitatively different contributions: a particle-dependent baseline that exists even for a spherical nucleus, and a deformation-induced modification generated during the QGP evolution. The normalization by the spherical configuration removes the former contribution and isolates the latter, allowing a direct comparison of how different hadronic species transport the deformation information.
	
	\begin{figure}[t!]
		\includegraphics[width=0.45\textwidth]{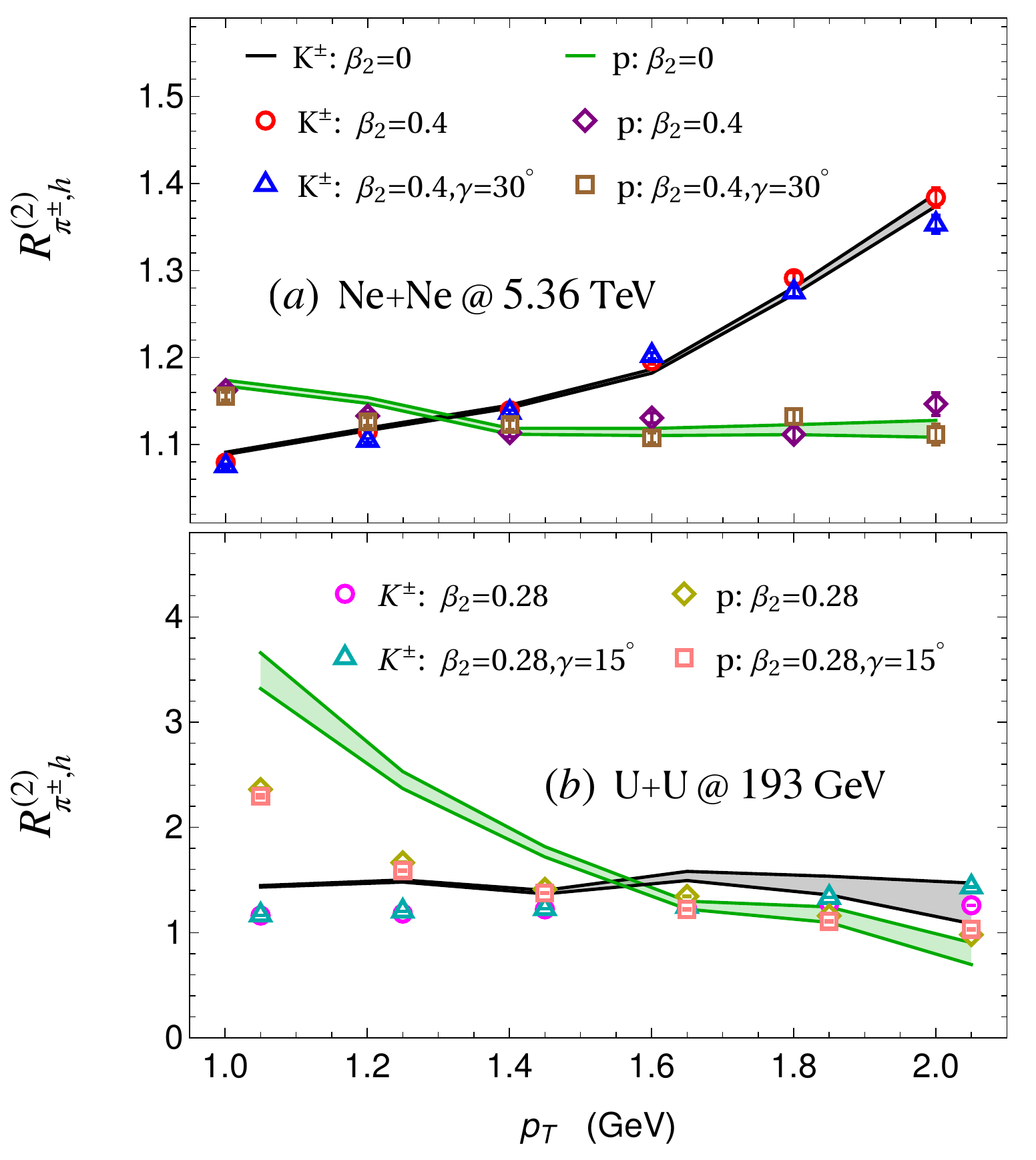}
		\caption{
			The same as Fig.~\ref{fig:4} but for $r_{2,2}(p_T)$. In Ne+Ne collisions, the different nuclear configurations remain nearly indistinguishable at low $p_T$ and separate only at high momentum, indicating that structural information is primarily encoded into the high-$p_T$ sector. In contrast, U+U collisions exhibit a stronger and earlier response to the initial deformation. The pion--kaon ratio increases with transverse momentum, whereas the pion--proton ratio decreases toward unity before saturating, producing a characteristic crossing between mesonic and baryonic response trajectories. The shift of the crossing region from Ne+Ne to U+U reflects collision-system and energy dependence of how the QGP transfers the initial nuclear geometry into identified-particle correlations.
		}
		\label{fig:4-4}
	\end{figure}
	Although $r_{2,2}(p_T)$ is experimentally more demanding than the
	two-particle coefficient $r_2(p_T)$ used in the main text, it provides
	an enhanced sensitivity to the deformation-dependent response.
	As illustrated in Fig.~\ref{fig:4-4}, the corresponding identified-particle ratios, using Eq.~(\ref{eq:four}),
	\begin{equation}\label{eqR}
		R_{\pi,h}^{(2)}(p_{\mathrm T})
		=
		\frac{r_{2,2}^{\pi}(p_{\mathrm T})}
		{r_{2,2}^{h}(p_{\mathrm T})}
		=
		\frac{\left\langle v_2^2(p_{\mathrm T})\right\rangle_{h}
			\left\langle V_2^2V_2^{*}(p_{\mathrm T})^2\right\rangle_{\pi}}
		{\left\langle v_2^2(p_{\mathrm T})\right\rangle_{\pi}
			\left\langle V_2^2V_2^{*}(p_{\mathrm T})^2\right\rangle_{h}}
	\end{equation}
	exhibit the same
	qualitative species dependence observed with $r_2(p_T)$, with
	deformation effects becoming more apparent in the relative responses
	of pions, kaons, and protons. In particular, the four-particle
	observable preserves the separation between mesonic and baryonic
	responses and reproduces the characteristic momentum dependence of
	the deformation-induced structures. This agreement provides a higher-order cross-check of the
	species-dependent deformation sensitivity identified with the two-particle decorrelation.

	For moderate deformations, the factorization-breaking coefficient of
	an identified particle $h$ can be expanded around the spherical
	configuration as
	\begin{align}
		r_h(\beta_2,\gamma)
		=
		r_{h}(0)
		+A_h(p_T)\beta_2^2
		+B_h(p_T)&\beta_2^3\cos(3\gamma)
		\nonumber\\&+C_h(p_T)\beta_2^4,
		\label{eq:expansion}
	\end{align}
	where $r_{h}(0)$ denotes the corresponding spherical result
	($r_{2;0}^h$ or $r_{2,2;0}^h$).  Defining the dimensionless
	single-particle response coefficients
	\begin{equation}
		a_h=\frac{A_h}{r_{h}(0)},\qquad
		b_h=\frac{B_h}{r_{h}(0)},\qquad
		c_h=\frac{C_h}{r_{h}(0)},
	\end{equation}
	gives
	\begin{equation}
		\frac{r_h(\beta_2,\gamma)}{r_{h}(0)}
		=
		1+a_h\beta_2^2
		+b_h\beta_2^3\cos(3\gamma)
		+c_h\beta_2^4.
		\label{eq:normalized-single}
	\end{equation}
	
	The normalized species response is a ratio of two such expansions,
	\begin{equation}
		\frac{R_{\pi,h}(\beta_2,\gamma)}
		{R_{\pi,h}(0)}
		=
		\frac{r_\pi/r_{\pi}(0)}{r_h/r_{h}(0)}.
	\end{equation}
	Expanding consistently through fourth order in $\beta_2$ yields
	\begin{align}
		\frac{R_{\pi,h}(\beta_2,\gamma)}
		{R_{\pi,h}(0)}
		=
		1
		&+\delta A_{\pi,h}(p_T)\beta_2^2
		+\delta B_{\pi,h}(p_T)\beta_2^3\cos(3\gamma)
		\nonumber\\
		&+\delta C_{\pi,h}^{\rm eff}(p_T)\beta_2^4,
		\label{eq:relative-expansion}
	\end{align}
	where
	\begin{align}
		\delta A_{\pi,h} &= a_\pi-a_h,\\
		\delta B_{\pi,h} &= b_\pi-b_h,\\
		\delta C_{\pi,h}^{\rm eff}
		&=c_\pi-c_h+a_h^2-a_\pi a_h
		\nonumber\\
		&=c_\pi-c_h-a_h\delta A_{\pi,h}.
		\label{eq:effective-quartic}
	\end{align}
	Thus, while the quadratic and cubic coefficients of the species ratio
	are simply differences of the corresponding single-particle response
	coefficients, its quartic coefficient contains an additional term
	generated by the expansion of the denominator. We denote this complete
	fourth-order coefficient by $\delta C_{\pi,h}^{\rm eff}$. It is this
	effective coefficient that enters the normalized species-response
	decompositions shown in Figs.~\ref{fig:7} and \ref{fig:14}.
	
	The coefficients $\delta A_{\pi,h}$, $\delta B_{\pi,h}$,  and $\delta C^{\rm eff}_{\pi,h}$ therefore do not represent the deformation sensitivity of an individual particle. Instead, they quantify the competition between two particle species in transporting the quadrupole and triaxial components of the initial nuclear geometry. Positive values indicate that pions preserve the corresponding deformation mode more efficiently than particle $h$, whereas negative values imply the opposite.
	
	\begin{figure*}[t!]
		\hspace*{-.7cm}\includegraphics[scale=.35]{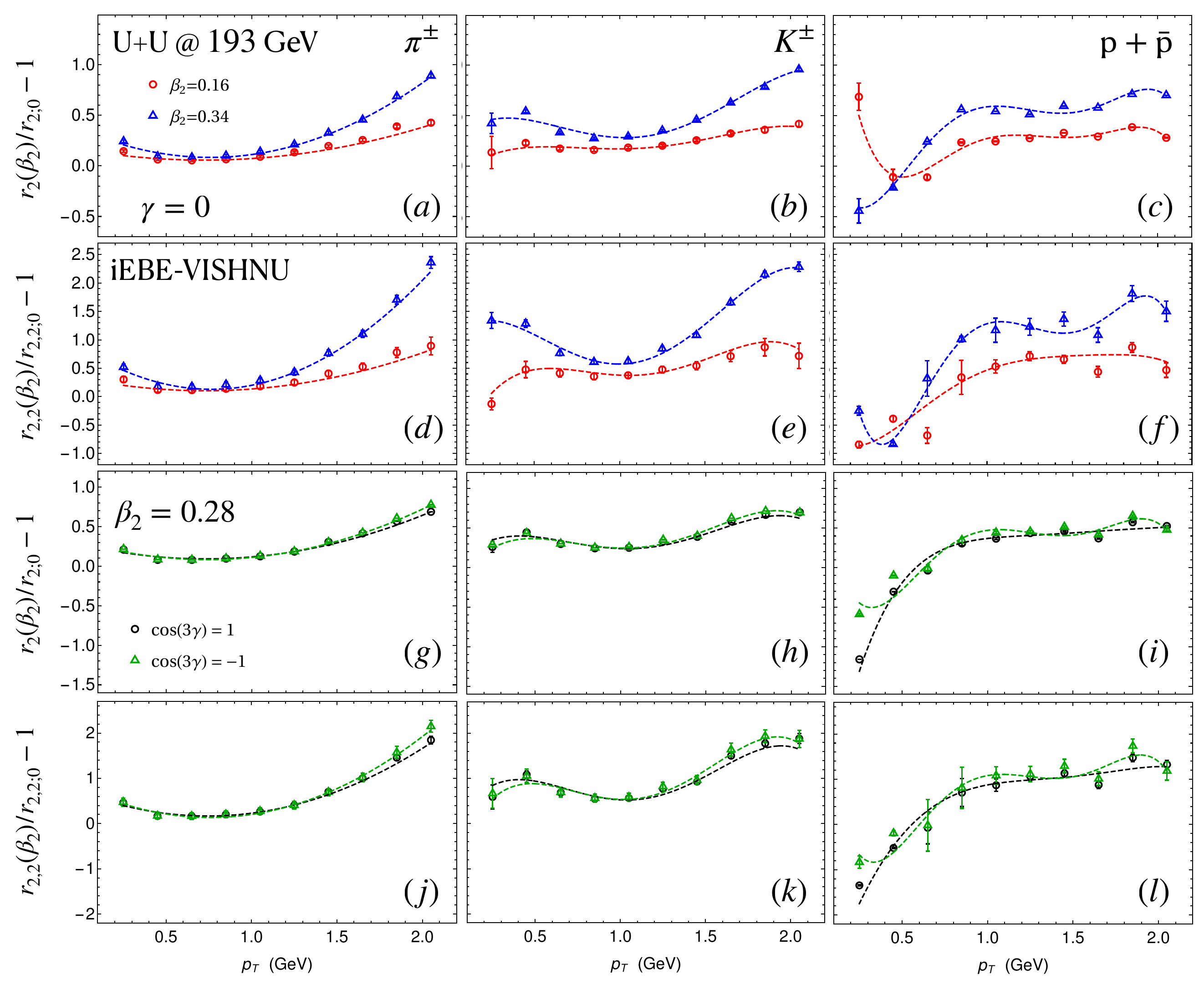}
		\caption{Momentum dependence of the deformation-induced response
			$r(\beta_2,\gamma)/r(0)-1$ for identified pions, kaons, and protons in
			central $^{238}$U+$^{238}$U collisions. Panels (a)--(c) show the
			two-particle decorrelation $r_2$ for $\beta_2=0.16$ and $0.34$ at
			$\gamma=0^\circ$, while panels (d)--(f) show the corresponding
			four-particle decorrelation $r_{2,2}$. Panels (g)--(i) show $r_2$ at
			fixed $\beta_2=0.28$ for $\cos(3\gamma)=1$ and $-1$, and panels
			(j)--(l) show the corresponding $r_{2,2}$ results. Dashed curves
			represent the deformation-expansion parameterizations described in
			the text. The deformation response exhibits a distinct
			particle- and momentum-dependent structure and is enhanced for the
			four-particle decorrelation.
		}
		\label{fig:6}
	\end{figure*}
	\begin{figure*}[t!]
		\includegraphics[scale=.40]{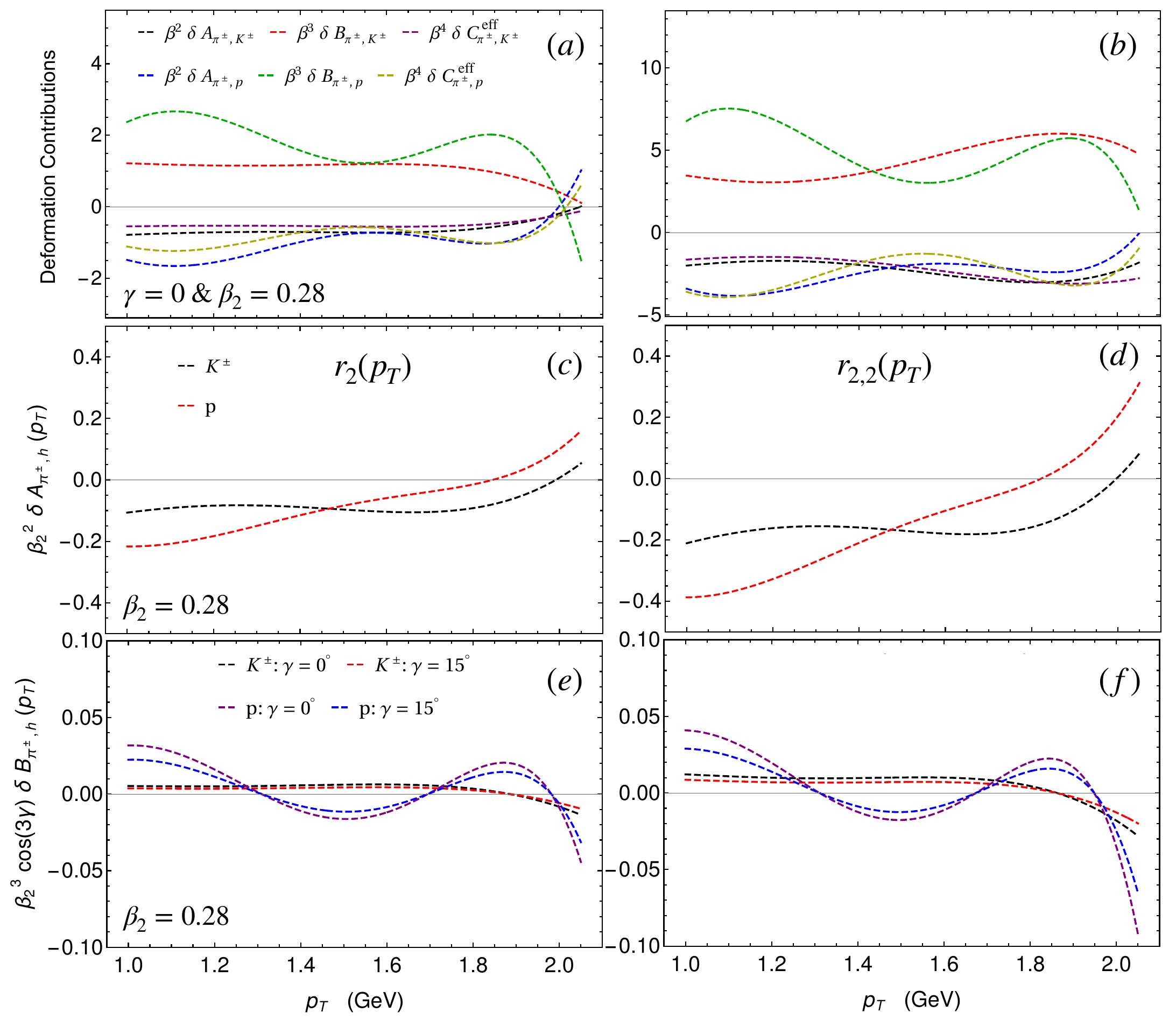}
		\caption{Decomposition of the deformation-dependent terms entering the
			normalized identified-particle response functions in central
			$^{238}$U+$^{238}$U collisions. Panels (a) and (b) show the individual
			$\beta_2^2$, $\beta_2^3$, and $\beta_2^4$ contributions to the
			pion--kaon and pion--proton responses at $\gamma=0^\circ$ for
			$r_2(p_T)$ and $r_{2,2}(p_T)$, respectively. Panels (c) and (d) show
			the momentum dependence of the quadratic contribution
			$\delta A_{\pi^\pm,h}(p_T)\beta_2^2$ for kaons and protons in $r_2$
			and $r_{2,2}$, respectively, at $\beta_2=0.28$. Panels (e) and (f)
			show the corresponding triaxial contribution
			$\delta B_{\pi^\pm,h}(p_T)\beta_2^3\cos(3\gamma)$ for
			$\gamma=0^\circ$ and $15^\circ$. The substantial cancellations among
			the deformation terms and their distinct momentum dependence for
			mesons and baryons account for the non-monotonic structures of the
			normalized responses in Fig.~\ref{fig:3}; the corresponding
			$r_{2,2}$ contributions exhibit the same structures with enhanced
			deformation sensitivity.}
		\label{fig:7}
	\end{figure*}
	\textit{Deformation expansion of the identified-particle response.--}
	To clarify the origin of the normalized response in
	Eq.~(\ref{eq:response}), we first examine the deformation dependence of
	the individual identified-particle decorrelations. Figure~\ref{fig:6}
	shows the expansion of Eq.~(\ref{eq:expansion})
	as a function of $p_T$ for pions, kaons, and protons in central
	$^{238}$U+$^{238}$U collisions, for both $r_2$ and $r_{2,2}$.
	Panels (a)--(f) compare two quadrupole deformations,
	$\beta_2=0.16$ and $0.34$, at $\gamma=0^\circ$, whereas
	panels (g)--(l) compare the axial and maximally triaxial limits,
	$\cos(3\gamma)=1$ and $-1$, at fixed $\beta_2=0.28$.
	
	For the quadrupole scan, the deformation response can be organized in
	powers of $\beta_2$ as
	\begin{equation}
		\frac{r(\beta_2)}{r(0)}-1
		=
		a_h(p_T)\beta_2^2
		+b_h(p_T)\beta_2^3
		+c_h(p_T)\beta_2^4.
		\label{eq:single_beta}
	\end{equation}
	The pion response varies smoothly with momentum and can be described
	with comparatively low-order $p_T$ dependence of the coefficients.
	The kaon and, particularly, proton responses exhibit more pronounced
	momentum structures and require higher-order $p_T$ parameterizations.
	This behavior is visible for both $r_2$ [panels (a)--(c)] and
	$r_{2,2}$ [panels (d)--(f)], with the four-particle decorrelation
	showing a substantially larger deformation response.
	
	At fixed $\beta_2$, the fitted quartic contribution is
	numerically negligible for the considered triaxiality scan.
	The leading deformation response can therefore be
	approximated as
	\begin{equation}
		\frac{r^h(\beta_2,\gamma)}{r^h(0)}-1
		\simeq a_h(p_{\mathrm T})\beta_2^2
		+b_h(p_{\mathrm T})\beta_2^3\cos(3\gamma).
		\label{eq:triaxial_response}
	\end{equation}
	The triaxiality dependence is thus governed, to this
	order, by the cubic contribution, while the quadratic
	term provides a $\gamma$-independent background.
	Panels (g)--(l) show that changing $\cos(3\gamma)$ from $1$ to $-1$
	produces only a modest modification of the pion and kaon responses,
	whereas the proton response exhibits a more pronounced dependence,
	particularly at low and intermediate $p_T$. The same qualitative
	species hierarchy is observed for $r_2$ and $r_{2,2}$, while the
	magnitude of the response is enhanced for $r_{2,2}$. These results
	demonstrate that the coefficients multiplying the allowed deformation
	structures have distinct momentum dependence for different identified
	particles.
	
	For the numerical parameterization, the pion coefficients are
	adequately described by quadratic polynomials in $p_T$. For the
	$\beta_2$ scan, the stronger momentum dependence of the kaon and proton
	responses requires terms up to fourth order in $p_T$, including the
	$\beta_2^4$ contribution. For the triaxial scan, fourth-order
	polynomials are sufficient for kaons, whereas the proton response
	requires terms up to fifth order. These polynomial orders should be
	understood as a convenient parameterization of the simulated
	$p_T$ dependence rather than as additional deformation modes.
	
	It is important that $b_h(p_T)$ is an intrinsic response coefficient and
	does not vanish for an axially symmetric nucleus. At $\gamma=0^\circ$,
	$\cos(3\gamma)=1$, and the cubic contribution is therefore
	$b_h(p_T)\beta_2^3$. A scan in $\beta_2$ consequently contains both
	quadratic and cubic deformation responses (as well as the quartic
	contribution where required), whereas varying $\gamma$ at fixed
	$\beta_2$ modulates the cubic term through $\cos(3\gamma)$. This
	separation allows the quadrupole-magnitude and triaxial contributions
	to be traced independently in the identified-particle response.
	
	Figure~\ref{fig:7} decomposes the deformation-dependent response into
	its individual contributions and provides the theoretical basis for
	Eq.~(\ref{eq:response}). For $\gamma=0^\circ$, the normalized response
	contains the three deformation terms
	$\delta A_{\pi,h}\beta_2^2$,
	$\delta B_{\pi,h}\beta_2^3$, and
	$\delta C_{\pi,h}^{\rm eff}\beta_2^4$. We emphasize that $c_h$ and $\delta C_{\pi,h}^{\rm eff}$ have
	different meanings: the former is the intrinsic quartic coefficient
	of the single-species response, whereas the latter is the complete
	quartic coefficient generated after forming the normalized ratio of
	two species.
	As shown in Figs.~\ref{fig:7}(a) and (b), these contributions
	are individually sizable and undergo substantial
	cancellations over the considered momentum range.
	For the four-particle construction in Fig.~\ref{fig:7}(b), at
	$\beta_2=0.28$ and $p_{\mathrm T}=1$~GeV, the pion--kaon
	contributions are approximately $-2.00$, $+3.47$,
	and $-1.64$, respectively, while the corresponding
	pion--proton contributions are approximately $-3.40$,
	$+6.78$, and $-3.59$. Their cancellations yield the
	much smaller net deformation responses shown in
	Figs.~~\ref{fig:3}(b) and (f).
	As $p_T$ increases, the balance among the three terms
	changes, producing the rise of the normalized response and its
	crossing of unity. The quartic contribution is therefore necessary to
	reproduce the full momentum dependence, particularly in the proton
	channel. The same decomposition is found for $r_2$ and $r_{2,2}$,
	with the latter exhibiting an enhanced deformation sensitivity.
	
	For fixed $\beta_2$, the $\gamma$ dependence is governed by the
	triaxial contribution
	$\delta B_{\pi,h}\beta_2^3\cos(3\gamma)$, whereas the quadratic
	contribution is independent of $\gamma$. Figures~\ref{fig:7}(c) and
	(d) show that the quadratic contribution has a distinct momentum
	dependence for kaons and protons, while the triaxial contribution is
	relatively small for kaons but becomes substantially larger for
	protons and changes sign with momentum. The resulting interplay
	explains the stronger separation of the triaxial configurations in the
	proton response at higher $p_T$. Figures~\ref{fig:7}(e) and (f) show the
	corresponding behavior for the four-particle decorrelation, where the
	same deformation structures are enhanced. Thus, the particle-dependent
	response arises from the momentum-dependent competition among multiple
	deformation contributions, providing the basis for the complementary
	mesonic and baryonic responses used in the main analysis.
	\begin{figure}[t!]
		\hspace*{-.4cm}\includegraphics[scale=.27]{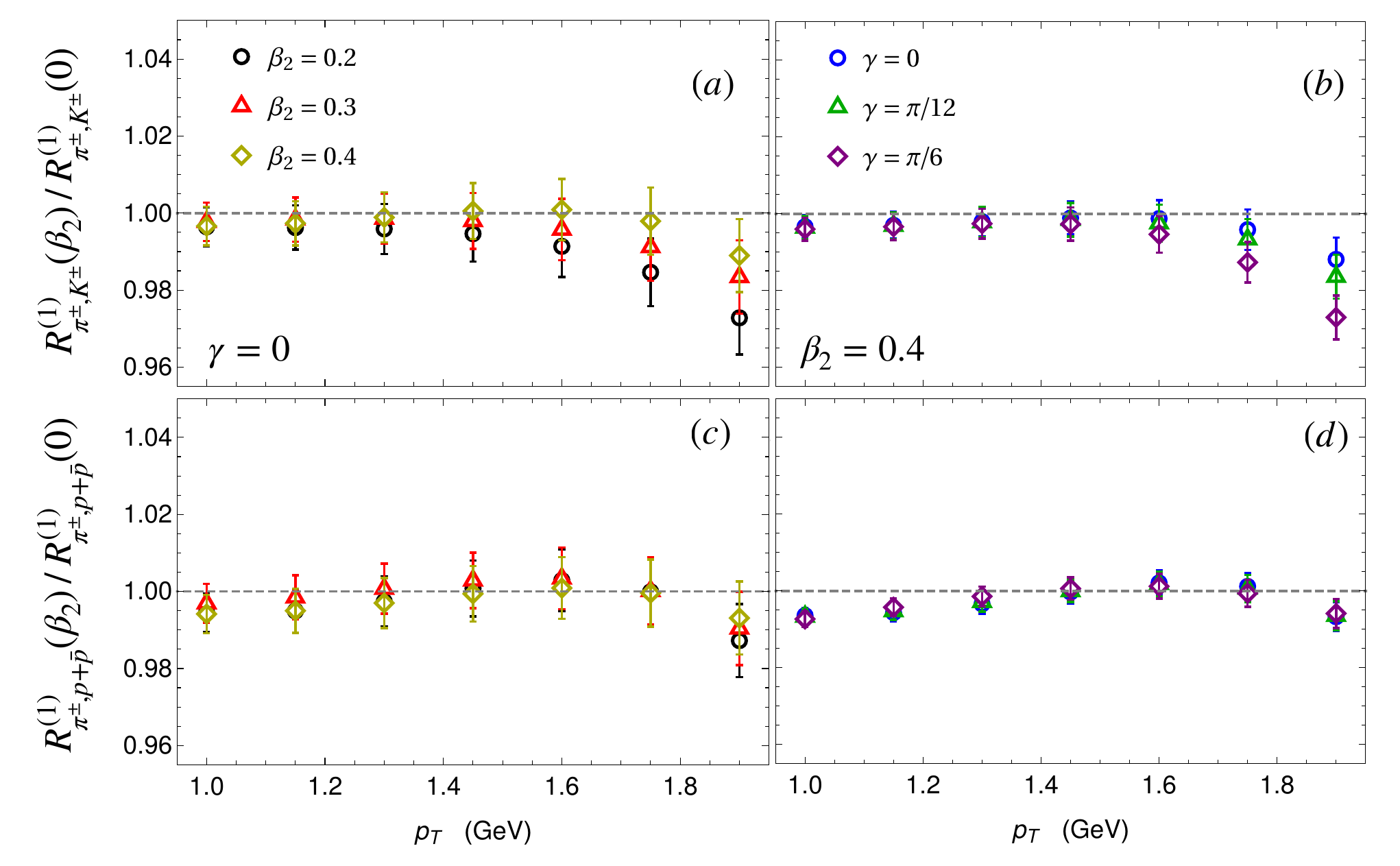}
		\caption{Normalized identified-particle response functions constructed from
			$r_2(p_T)$ in central $^{20}$Ne+$^{20}$Ne collisions at
			$\sqrt{s_{\rm NN}}=5.36$~TeV. Panels (a) and (c) show the
			pion--kaon and pion--proton responses, respectively, for different
			$\beta_2$ at $\gamma=0^\circ$, while panels (b) and (d) show the
			corresponding triaxial dependence at fixed $\beta_2=0.4$.
			Although the deformation-induced deviations from unity are only at the percent level, their distinct momentum and species dependences reveal the relative corrections to the direct species ratios that determine the crossing coordinates in Fig.~\ref{fig:5}.}
		\label{fig:12}
	\end{figure}
	\subsection{Origin of the identified-particle response in
		$^{20}$Ne+$^{20}$Ne}
	
\begin{figure*}[t!]
		\hspace*{-.4cm}\includegraphics[scale=.39]{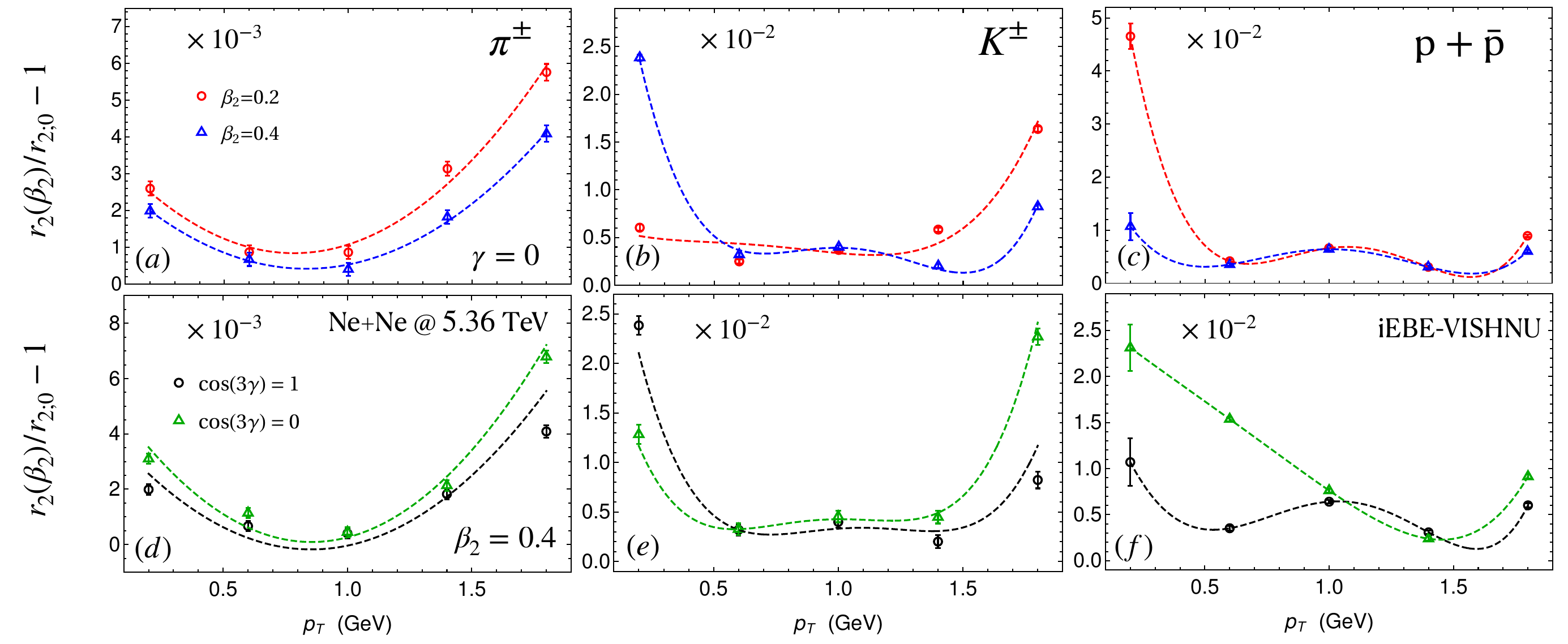}
		\caption{Deformation-induced fractional modification of the individual
			identified-particle two-particle flow decorrelation in central
			$^{20}$Ne+$^{20}$Ne collisions at
			$\sqrt{s_{\rm NN}}=5.36$~TeV,
			$r_2(\beta_2,\gamma)/r_{2,0}-1$.
			Panels (a)--(c) show the quadrupole response of pions, kaons, and
			protons, respectively, at $\gamma=0^\circ$, while panels (d)--(f)
			show the triaxial response at fixed $\beta_2=0.4$.
			Notice the different vertical scale factors: the pion modification
			is at the $10^{-3}$ level, whereas the kaon and proton responses
			reach the $10^{-2}$ level.
			The distinct momentum dependence among particle species generates
			the small non-universal component that survives in the relative
			responses of Fig.~\ref{fig:12}.}
		\label{fig:13}
	\end{figure*}

\begin{figure}[t!]
		\hspace*{-.4cm}\includegraphics[scale=.39]{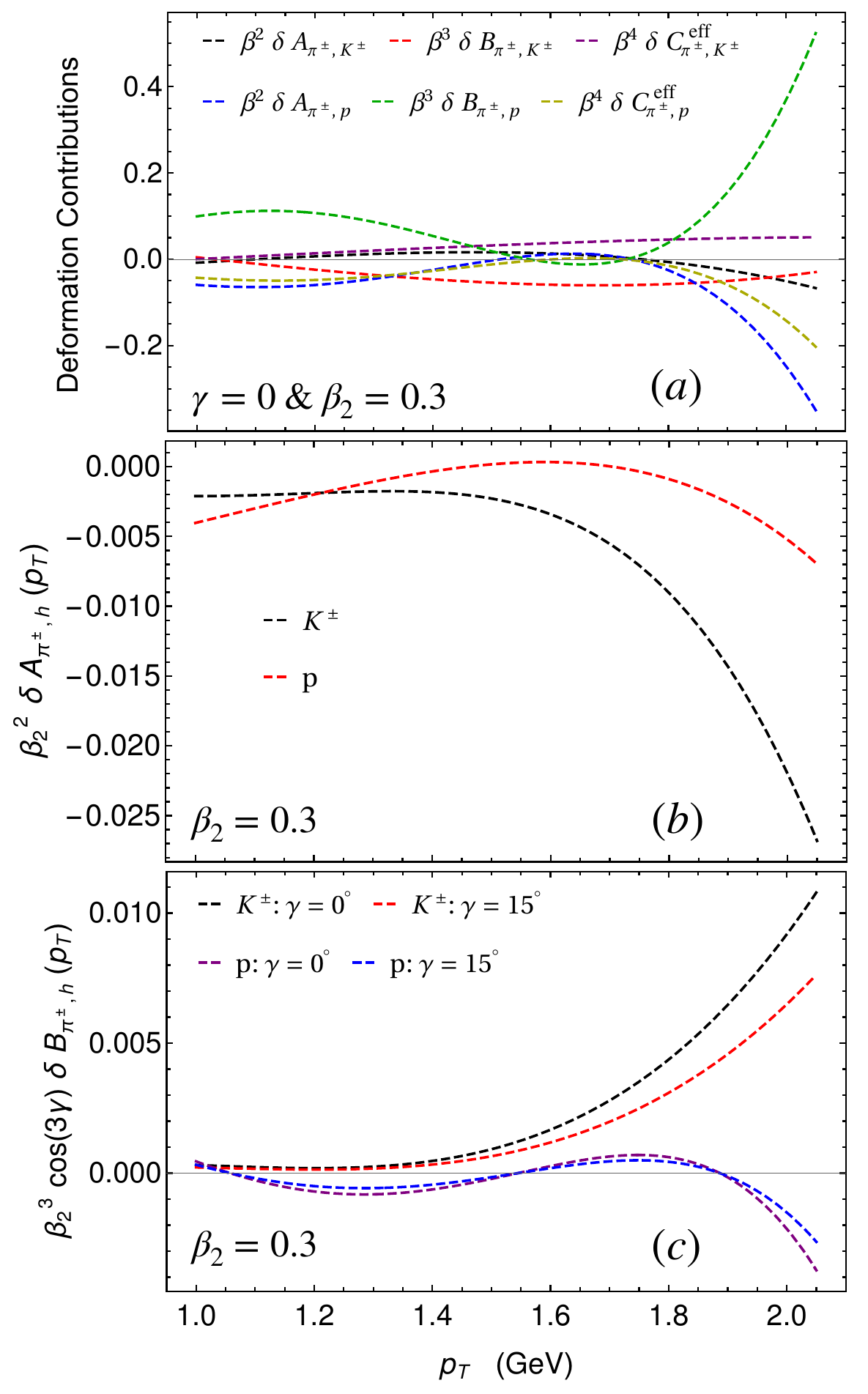}
		\caption{Decomposition of the deformation-dependent normalized particle
			responses in central $^{20}$Ne+$^{20}$Ne collisions at
			$\sqrt{s_{\rm NN}}=5.36$~TeV.
			Panel (a) shows the individual
			$\beta_2^2\delta A_{\pi,h}$,
			$\beta_2^3\delta B_{\pi,h}$, and
			$\beta_2^4\delta C_{\pi,h}^{\rm eff}$ contributions for the pion--kaon and
			pion--proton channels at $\beta_2=0.3$ and $\gamma=0^\circ$.
			Panel (b) shows the quadratic contribution
			$\beta_2^2\delta A_{\pi,h}(p_T)$, while panel (c) shows the triaxial
			term
			$\beta_2^3\cos(3\gamma)\delta B_{\pi,h}(p_T)$
			for $\gamma=0^\circ$ and $15^\circ$.
			The small net particle response results from substantial
			momentum-dependent cancellations among individually larger deformation
			terms. Their different signs and momentum dependences for kaons and
			protons provide the microscopic origin of the deformation-dependent
			crossing coordinates in Fig.~\ref{fig:5}.}
		\label{fig:14}
	\end{figure}

	The compact crossing-coordinate map obtained for
	$^{20}$Ne+$^{20}$Ne collisions deserves further examination.
	As shown in the main text, the deformation dependence of the
	individual identified-particle decorrelations is considerably weaker
	than in $^{238}$U+$^{238}$U, while the crossing coordinates nevertheless
	remain sensitive to both $\beta_2$ and $\gamma$.
	Figures~\ref{fig:12}--\ref{fig:14} clarify how these two observations
	are reconciled.  The essential point is that, in the light system,
	nuclear deformation produces relatively small but strongly
	momentum- and species-dependent modifications of the individual
	decorrelations.  Forming ratios between particle species suppresses their common response while retaining the differential deformation dependence, whose momentum evolution modifies the direct species ratios and consequently shifts their crossing coordinates.
	
	Figure~\ref{fig:12} shows the normalized identified-particle response
	functions
	\begin{equation*}
		\frac{R_{\pi,h}^{(1)}(\beta_2,\gamma)}
		{R_{\pi,h}^{(1)}(0)},
		\qquad h=K,p ,
	\end{equation*}
	constructed from the two-particle decorrelation $r_2(p_T)$.
	For both the quadrupole and triaxial scans, these ratios remain close
	to unity over most of the displayed momentum interval.  At
	$\gamma=0^\circ$, the pion--kaon response differs from unity by less
	than about $1\%$ up to $p_T\simeq1.7$~GeV.  The separation becomes
	more visible toward the upper end of the momentum range, where the
	response at $p_T\simeq1.9$~GeV is approximately $0.973$, $0.984$,
	and $0.988$ for $\beta_2=0.2$, $0.3$, and $0.4$, respectively.
	The pion--proton response displays an even weaker dependence on
	$\beta_2$ over the same interval.
	
	A similarly small but systematic structure is generated by
	triaxiality at fixed $\beta_2=0.4$.  The pion--kaon curves are nearly
	degenerate at low and intermediate $p_T$ but progressively separate
	toward high momentum.  Around $p_T\simeq1.9$~GeV, the response
	decreases from approximately $0.987$ for $\gamma=0^\circ$ to
	$\sim0.983$ for $\gamma=15^\circ$ and $\sim0.973$ for
	$\gamma=30^\circ$.  By comparison, the pion--proton response remains
	much closer to unity and exhibits a substantially weaker triaxial
	splitting.  The deformation signal in $R_{\pi,h}^{(1)}$ is therefore
	small in magnitude but not featureless: its sign, momentum dependence,
	and deformation dependence differ between the pion--kaon and
	pion--proton channels.  It is precisely this differential structure
	that controls the position at which the two response functions
	intersect and hence the crossing coordinates shown in
	Fig.~\ref{fig:5}.
	
	To understand the origin of this residual response, Fig.~\ref{fig:13}
	shows the deformation-induced modification of each identified-particle
	decorrelation before taking ratios between species,
	\begin{equation*}
		\frac{r_2^h(\beta_2,\gamma;p_T)}
		{r_{2,0}^h(p_T)}-1 ,
		\qquad h=\pi,K,p .
		\label{eq:Ne-individual-response}
	\end{equation*}
	The scale factors displayed in the panels are important:
	the pion modifications are at the $10^{-3}$ level, whereas the kaon
	and proton modifications can reach the $10^{-2}$ level.  Thus,
	Fig.~\ref{fig:13} is quantitatively consistent with the weak
	deformation dependence of the individual $r_2(p_T)$ distributions
	shown in Fig.~\ref{fig:2}.

	For the quadrupole scan at $\gamma=0^\circ$, the pion modification
	remains below approximately $6\times10^{-3}$ over the displayed
	momentum range.  It exhibits a shallow minimum at intermediate
	$p_T$ and increases toward high momentum, reaching about
	$5.8\times10^{-3}$ for $\beta_2=0.2$ and
	$4.1\times10^{-3}$ for $\beta_2=0.4$ at
	$p_T\simeq1.8$~GeV.  The corresponding kaon response is larger and
	more strongly nonmonotonic.  For $\beta_2=0.4$, for example, it is
	about $2.4\times10^{-2}$ at low $p_T$, decreases to a few
	$10^{-3}$ around $p_T\simeq0.6$--$1.4$~GeV, and rises again to
	approximately $8\times10^{-3}$ at the upper end of the range.
	The proton channel shows an even stronger low-$p_T$ modification:
	for $\beta_2=0.2$ it reaches approximately
	$4.6\times10^{-2}$ at the lowest momentum before rapidly decreasing
	to the few-$10^{-3}$ level at intermediate $p_T$.

	These results expose an important feature of the light-ion response.
	The weak deformation dependence seen in the individual
	$r_2(p_T)$ distributions does not mean that all identified particles
	respond identically.  Rather, the fractional corrections are
	different for pions, kaons, and protons and evolve differently with
	$p_T$.  Since
	$R_{\pi,h}^{(1)}=r_2^\pi/r_2^h$, contributions that modify the two
	species similarly are suppressed in the ratio, whereas differences
	between their deformation responses survive.  The near-unity values
	in Fig.~\ref{fig:12} therefore quantify the small
	\emph{non-universal} component of the deformation response rather
	than the absolute deformation effect on either species separately.

	The lower panels of Fig.~\ref{fig:13} demonstrate that the same
	mechanism operates for triaxiality.  At fixed $\beta_2=0.4$, changing
	from $\cos(3\gamma)=1$ to $\cos(3\gamma)=0$ modifies both the
	magnitude and momentum dependence of the individual decorrelations.
	The effect remains at the sub-percent level for pions but reaches the
	percent level for kaons and protons.  In particular, the kaon
	modification changes appreciably both at low and high $p_T$, while
	the proton response shows a pronounced triaxial splitting at low
	momentum.  Consequently, even though the absolute deformation
	corrections remain modest, their different momentum evolution among
	particle species generates the differential response observed in
	Fig.~\ref{fig:12}.
	
	The origin of its nontrivial momentum dependence can be examined more
	directly through the deformation expansion shown in
	Fig.~\ref{fig:14}. 
	Panel~(a) displays the three contributions separately for
	$\beta_2=0.3$.  A striking feature is that the individual terms can
	be considerably larger than the final response because they carry
	different signs and momentum dependences.  This is particularly clear
	in the pion--proton channel at high $p_T$: the cubic contribution
	grows rapidly and reaches approximately $+0.5$ near
	$p_T\simeq2.05$~GeV, while the quadratic and quartic contributions
	become negative, reaching roughly $-0.35$ and $-0.2$, respectively.
	The resulting response is therefore controlled by a substantial
	cancellation among different orders in the deformation expansion.
	The pion--kaon channel exhibits an analogous competition, although
	with a different balance among the three terms.
	
	Panels~(b) and (c) make the species dependence of this cancellation
	particularly transparent.  The quadratic contribution to the
	pion--kaon response becomes increasingly negative toward high
	$p_T$, reaching approximately $-2.5\times10^{-2}$ at
	$p_T\simeq2.05$~GeV, whereas the corresponding pion--proton
	contribution remains considerably smaller over most of the momentum
	range.  The hierarchy persists in the cubic term.  In the pion--kaon
	channel it becomes positive and rises rapidly above
	$p_T\simeq1.5$~GeV, reaching approximately
	$1.1\times10^{-2}$ for $\gamma=0^\circ$ near the upper end of the
	range, while the pion--proton contribution remains close to zero over
	much of the interval and changes sign at high $p_T$.
	
	Interestingly, this hierarchy is reversed relative to the U+U
	results in Fig.~\ref{fig:7}.  There, the deformation-dependent
	relative response is more pronounced in the pion--proton channel,
	whereas in Ne+Ne at $\sqrt{s_{\rm NN}}=5.36$~TeV the pion--kaon
	channel carries the larger quadratic and triaxial contributions.
	The species that most strongly carries the residual deformation
	information is therefore not universal.  Rather, the relative
	mesonic and baryonic responses depend on the dynamical scales of the
	collision, including the system size and collision energy.  This
	observation is particularly relevant for the crossing-coordinate
	construction, since $P_{\rm cross}$ is determined by the competition
	between precisely these two particle-response channels.
	
	\begin{figure}[t!]
		\hspace*{-.2cm}\includegraphics[scale=.35]{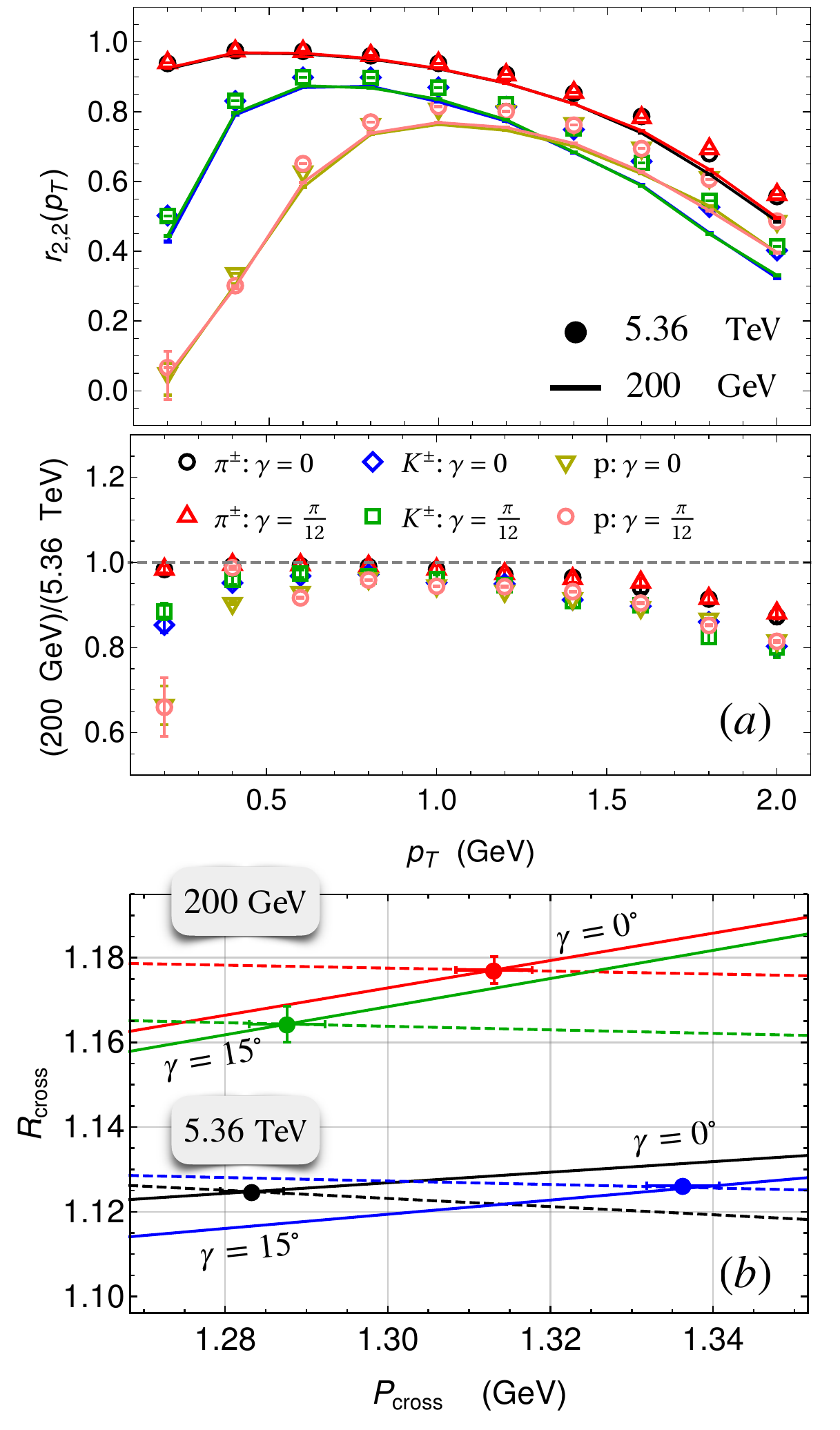}
		\caption{Collision-energy dependence of the identified-particle decorrelations
			and crossing coordinates in central $^{20}$Ne+$^{20}$Ne collisions.
			Panel (a) compares $r_{2,2}(p_T)$ at
			$\sqrt{s_{\rm NN}}=5.36$~TeV (symbols) and $200$~GeV (curves) for
			$(\beta_2,\gamma)=(0.4,0^\circ)$ and $(0.4,15^\circ)$; the lower
			panel shows their ratio. Panel (b) shows the corresponding crossing
			coordinates. Lowering the collision energy shifts the crossing map and
			reverses the direction of the triaxial displacement in
			$P_{\rm cross}$, reflecting the energy dependence of the relative
			pion--kaon and pion--proton responses near their intersection.
			Nevertheless, the two nuclear configurations remain separated,
			demonstrating that the sensitivity of $C_{\rm cross}$ to intrinsic
			nuclear geometry persists across collision energies.}
		\label{fig:15}
	\end{figure}

	The energy dependence of this hierarchy provides a natural
	connection to the study below.  As demonstrated in
	Fig.~\ref{fig:15}, reducing the Ne+Ne collision energy from
	$5.36$~TeV to $200$~GeV changes the relative slopes of
	$R_{\pi,K}$ and $R_{\pi,p}$ and consequently changes the direction
	in which the crossing coordinate moves with triaxiality.  At
	$5.36$~TeV, increasing $\gamma$ from $0^\circ$ to $15^\circ$
	shifts $P_{\rm cross}$ from $1.2833$ to $1.3213$~GeV, whereas at
	$200$~GeV the same variation shifts it oppositely, from $1.3130$
	to $1.2876$~GeV.  Correspondingly, $R_{\rm cross}$ changes from
	$1.1248$ to $1.1262$ at $5.36$~TeV and from $1.1770$ to $1.1643$
	at $200$~GeV.  Thus, the kaon-dominated deformation response seen
	in Fig.~\ref{fig:14} is part of a broader energy-dependent
	reorganization of the identified-particle response, rather than a
	generic property of light nuclei.
	
	Figures~\ref{fig:12}--\ref{fig:14} therefore provide the microscopic
	origin of the compact Ne+Ne crossing map.  Nuclear deformation
	produces modest but species-dependent modifications of the individual
	decorrelations, Fig.~\ref{fig:13}; their common component is largely
	suppressed in the particle ratios, leaving the small differential
	response of Fig.~\ref{fig:12}.  Figure~\ref{fig:14} shows that this
	residual response results from a momentum-dependent competition among
	the $\beta_2^2$, $\beta_2^3\cos(3\gamma)$, and $\beta_2^4$
	contributions, with the pion--kaon channel playing the dominant role
	in Ne+Ne at $5.36$~TeV.  Since the balance between the pion--kaon
	and pion--proton channels evolves with the collision scale, their
	intersection provides sensitivity not only to the intrinsic geometry
	but also to how that geometry is transported through the QGP.
	The energy-dependence study in Fig.~\ref{fig:15} tests this
	interpretation directly.

	\subsection*{Collision-energy dependence of the crossing coordinates}

	The preceding analysis shows that the crossing coordinates originate
	from a momentum-dependent competition among the identified-particle
	responses. It is therefore natural to ask whether this mapping is
	modified by the collision energy. Figure~\ref{fig:15} compares
	$^{20}$Ne+$^{20}$Ne collisions at $\sqrt{s_{\rm NN}}=5.36$~TeV and
	$200$~GeV for $(\beta_2,\gamma)=(0.4,0^\circ)$ and
	$(0.4,15^\circ)$.
	
	As shown in Fig.~\ref{fig:15}(a), lowering the collision energy changes
	both the magnitude and the $p_T$ dependence of the identified-particle
	decorrelations. These modifications are not identical for pions,
	kaons, and protons and consequently change the relative slopes of
	$R_{\pi,K}(p_T)$ and $R_{\pi,p}(p_T)$ in the momentum interval where
	the two responses intersect. The crossing coordinates therefore probe
	not only the magnitude of the deformation response, but also its
	energy-dependent momentum evolution.
	
	This behavior is evident in Fig.~\ref{fig:15}(b). At
	$\sqrt{s_{\rm NN}}=5.36$~TeV, the axial and triaxial configurations
	give
	\begin{align}
		C_{\rm cross}(0.4,0^\circ)
		&=(1.2833,\;1.1248),\\
		C_{\rm cross}(0.4,15^\circ)
		&=(1.3213,\;1.1262),
	\end{align}
	respectively. Thus, triaxiality primarily shifts the crossing toward
	larger $P_{\rm cross}$, by $\Delta P_{\rm cross}\simeq+0.038$~GeV,
	while producing only a small change,
	$\Delta R_{\rm cross}\simeq+0.0014$, in its magnitude. At $200$~GeV, the corresponding coordinates become
	\begin{align}
		C_{\rm cross}(0.4,0^\circ)
		&=(1.3130,\;1.1770),\\
		C_{\rm cross}(0.4,15^\circ)
		&=(1.2876,\;1.1643).
	\end{align}
	The deformation trajectory changes qualitatively: increasing $\gamma$
	now shifts the crossing toward smaller momentum,
	$\Delta P_{\rm cross}\simeq-0.0254$~GeV, accompanied by a sizable
	decrease $\Delta R_{\rm cross}\simeq-0.0127$. The response to
	triaxiality therefore changes from an almost horizontal displacement
	in the crossing-coordinate plane at $5.36$~TeV to a correlated
	variation of both coordinates at $200$~GeV.
	
	The origin of this behavior can be traced to the energy dependence of
	the underlying identified-particle responses. As shown in
	Fig.~\ref{fig:15}(a), lowering the collision energy modifies both their
	magnitudes and their $p_T$ dependence, with different modifications
	for pions, kaons, and protons. Since $C_{\rm cross}$ is determined by
	the intersection of $R_{\pi,K}(p_T)$ and $R_{\pi,p}(p_T)$, its
	position depends sensitively on the relative values and slopes of
	these two responses near the crossing. The deformation contributions
	analyzed in Figs.~\ref{fig:12}--\ref{fig:14} are small and strongly
	momentum dependent in Ne+Ne; consequently, an energy-dependent change
	of their relative balance can alter not only the magnitude but also
	the direction of the deformation-induced displacement in the
	$(P_{\rm cross},R_{\rm cross})$ plane.
	
	Interestingly, the $200$~GeV trajectory is qualitatively similar to
	the behavior observed in $^{238}$U+$^{238}$U at $193$~GeV in
	Fig.~\ref{fig:5}, where increasing triaxiality at fixed $\beta_2$
	also drives the crossing toward smaller $P_{\rm cross}$ together with
	a change in $R_{\rm cross}$. This comparison indicates that the
	orientation of the Ne+Ne crossing map at $5.36$~TeV is not determined
	by nuclear size alone, but reflects the energy-dependent collective
	mapping of the initial geometry onto identified hadrons. Thus, while
	the absolute location and deformation trajectory of $C_{\rm cross}$
	depend on the collision energy, its sensitivity to the intrinsic
	nuclear geometry persists.

\end{document}